\documentclass[10pt,letterpaper,table]{article}
\usepackage[top=0.85in,left=2.75in,footskip=0.75in]{geometry}
\newsavebox{\measurebox}

\usepackage{amsmath,amssymb}

\usepackage{changepage}

\usepackage[utf8]{inputenc}

\usepackage{textcomp,marvosym}
\usepackage{svg}
\usepackage{bm}
\usepackage{xcolor}

\usepackage{subcaption}
\usepackage{mwe}

\usepackage{floatrow}
\newfloatcommand{capbtabbox}{table}[][\FBwidth]

\usepackage[right]{lineno}

\usepackage{microtype}
\DisableLigatures[f]{encoding = *, family = * }

\usepackage{array}

\newcolumntype{+}{!{\vrule width 2pt}}

\newlength\savedwidth

\raggedright
\usepackage[aboveskip=1pt,labelfont=bf,labelsep=period,justification=raggedright,singlelinecheck=off]{caption}

\usepackage{cite}

\usepackage{lastpage,fancyhdr,graphicx}
\usepackage{epstopdf}
\usepackage{comment}

\fancyheadoffset[L]{2.25in}
\fancyfootoffset[L]{2.25in}
\usepackage{blindtext}
\usepackage{tabularx}
\usepackage{graphicx}
\usepackage{subcaption}
\usepackage{physics}
\usepackage{bm}
\usepackage[labelformat=parens,labelsep=quad,skip=3pt]{caption}
\usepackage{mathtools}
\usepackage{tikz}
\usetikzlibrary{arrows.meta,positioning}
\usepackage{amsfonts}
\usepackage[utf8]{inputenc} 

\usepackage{textcomp}
\usepackage{float}
\usepackage[version=4]{mhchem} 
\usepackage{booktabs}
\usepackage[flushleft]{threeparttable}
\usepackage{array,booktabs,makecell}
\usepackage[font=small]{caption}
\newlength{\cewidth}

\newcommand{\overbar}[1]{\mkern 1.5mu\overline{\mkern-1.5mu#1\mkern-1.5mu}\mkern 1.5mu}
\usepackage{comment}
\usepackage{setspace} 

\begin{document}
\vspace*{0.2in}

\begin{flushleft}
{\Large
\textbf\newline{The Role of RNA Condensation in Reducing Gene Expression Noise} 
}
\newline
\\
Alex Mayer$^1$, Grace McLaughlin$^{2}$, Sierra Cole$^{3}$, Amy Gladfelter$^{2}$, Marcus Roper$^{1,4}$
\\
\bigskip
\textbf{1} Department of Mathematics, UCLA, Los Angeles, California, United States of America
\\
\textbf{2} Department of Biology, UNC, Chapel Hill, North Carolina, United States of America
\\
\textbf{3} Department of Biochemistry and Biophysics, UNC, Chapel Hill, North Carolina, United States of America
\\
\textbf{4} Department of Computational Medicine, UCLA, Los Angeles, California, United States of America

\bigskip



\end{flushleft}
\section*{Abstract}

Biomolecular condensates have been shown to play a fundamental role in localizing biochemistry in a cell. RNA is a common constituent of condensates, and can determine their biophysical properties. Functions of biomolecular condensates are varied including activating, inhibiting, and localizing reactions. Recent theoretical work has shown that  the phase separation of proteins into droplets can diminish cell to cell variability in protein abundance. However, the extent to which phase separation involving mRNAs may also buffer noise has yet to be explored. In this paper, we introduce a phenomenological model for the phase separation of mRNAs into RNP condensates, and quantify noise suppression as a function of gene expression kinetic parameters. Through stochastic simulations, we highlight the ability for condensates formed from just a handful of mRNAs to regulate the abundance and suppress the fluctuations of proteins. We place particular emphasis on how this mechanism can facilitate efficient transcription by reducing noise even in the situation of infrequent bursts of transcription by exploiting the physics of a concentration-dependent, deterministic phase separation threshold. We investigate two biologically relevant models in which phase separation acts to either "buffer" noise by storing mRNA in inert droplets, or "filter" phase separated mRNAs by accelerating their decay, and quantify expression noise as a function of kinetic parameters. In either case the most efficient expression occurs when bursts produce mRNAs close the phase separation threshold, which we find to be broadly consistent with observations of an RNP-droplet forming cyclin in multinucleate \textit{Ashbya gossypii} cells. We finally consider the contribution of noise in the phase separation threshold, and show that protein copy number noise can be efficiently suppressed by phase separation threshold fluctuations in certain conditions. 

\section*{Author Summary}

Due to the inherent noise of gene expression, the quantity of any protein contained within a cell may fluctuate over time. In particular cells must trade off the efficiency of translating a large number of proteins from each mRNA with the amplification of Poisson noise produced when only handfuls of mRNA transcripts are expressed. 
Phase separation of proteins into biomolecular condensates is already known to be able to reduce fluctuations in soluble protein. However, many condensates are composed of RNA and the contribution of these droplets to buffer RNA and protein fluctuations has been less investigated. Here we develop a mathematical model to show that RNP droplets may serve to reduce variability in the levels of protein encoded by the mRNAs, through two possible mechanisms, which we call buffering and filtering. Our results suggest that phase separation is particularly favored when mRNAs are synthesized in infrequent, thereby noisy bursts that are tuned to the concentration threshold for phase separation, increasing the efficiency of protein expression. Analysis of smFISH data from an RNP forming system supports this hypothesis, and compel further experiments to explore the link between transcriptional bursts and RNA phase separation.  


\section*{Introduction}

Gene expression is a noisy process, ensuring that even genetically identical cells receiving common cues from their environment may exhibit a range of protein copy numbers. In some cases, resulting cell to cell variability might be useful \cite{maheshri2007living} \cite{raser2005noise}. However, in stable environments, cells generally benefit from consistent expression of proteins, and high gene expression noise may impair cellular function \cite{charlebois2015effect} \cite{wang2011impact} \cite{bahar2006increased}. Low mRNA copy numbers are a major contributor to protein copy noise since translation to proteins amplifies small absolute variations in mRNA copy number \cite{buccitelli2020mrnas}. Yet, mRNA copy numbers are often low - at 10 mRNAs or fewer per cell, across the majority of the genome in \textit{Saccharomyces cerevisiae} \cite{miura2008absolute}, and transcription rates (number of mRNAs per gene per unit time) are much smaller than translation rates (number of proteins per mRNA copy per unit time) across the genomes of yeast, mice, humans and E. coli in their fast growing phases \cite{hausser2019central}. Hausser et al \cite{hausser2019central} argue that energetics constrain transcription rates;  the total energetic cost of translation is invariant if protein copy numbers are held fixed, but the energetic burden of transcription, although relatively smaller, is reduced if mRNA copy numbers are kept small. Economical transcription may also lessen transcriptional interference: the negative interactions of different transcriptional activities due, for example, to elongating RNA polymerases obstructing each other, repressors bound to one operon overlapping with a second operon \cite{shearwin2005transcriptional}, or from genome conformational changes that expose one operon, but mask another \cite{kim2019mechanisms}. 
\newline 
\newline
In higher eukaryotes, the primary mode of transcription is through bursting \cite{nicolas2017shapes}.  During bursty transcription, mRNAs are synthesized in pulses. These pulses are generally assumed to constitute a Poisson point process, though evidence suggests that the arrival process may be more complex. \cite{kumar2015transcriptional}.
Bursts are thought to arise as a result of the reversible interactions of the pre-initiation complex (PIC) with a gene's promoter. 
When the PIC is bound, the RNA polymerase's affinity for binding to the transcription start site is dramaticallly 
increased, and multiple transcription events can occur while the gene is in this ``on" state \cite{boeger2015structural}. 
Since chromatin remodeling is required for gene activation, 
there is a fundamental cost to fast switching between promoter states \cite{huang2015fundamental}. Efficiency can thus be achieved by minimizing the frequency of gene activity, 
and allowing intense bursts when in the ``on" state to meet the required mean mRNA abundance.  However bursty transcription introduces additional noise into expression. In the context of infrequent, intense bursts, mRNA variability will drastically increase, which is especially pronounced in systems with small mean mRNA populations, resulting in large steady state fluctuations in protein.
\newline
\newline
For many genes, protein abundance must be tightly regulated for proper function of the cell \cite{barkai2000circadian} \cite{lehner2008selection} \cite{wang2011impact}, and there exist many regulatory mechanisms in gene expression that can mitigate fluctuations in protein abundance \cite{tan2021quantitative}, \cite{singh2011negative}. Negative feedback has been observed at each level of gene expression, resulting in gene networks that can be analyzed for their ability to control noise in expression \cite{singh2011negative}. These networks can theoretically drive expression noise below Poisson levels, but can introduce substantial deficits in the cell's energy economy \cite{stoeger2016passive}. For example, Lestas et al. \cite{lestas2010fundamental} find that in systems with nonlinear real-time feedback control, signal molecules must be synthesized at rates far exceeding those of the target molecule in order to meaningfully suppress noise. In contrast, cellular compartmentalization may an energetically efficient method for filtering expression noise if the proteins that form the compartments are long-lived  \cite{stoeger2016passive}. In particular, phase separation has been theorized to play a role in post-translational regulation of genes. Klosin et. al \cite{klosin2020phase} provide  theoretical and experimental evidence of this idea; showing that concentration-dependent phase separation of proteins can drive protein fluctuations to the minimum Poisson noise limit of the network. Deviri and Safran  \cite{deviri2021physical} extended this theory to multi-component phase separation, and derived criteria on their phase diagrams under which concentration buffering may occur. They hypothesized that in genes that are sensitive to noise, selective pressures may act to optimize concentration buffering, though the extent to which noise is suppressed in these systems remains less clear.
\newline
\newline
One such class of multi-component condensates are ribonucleoprotein (RNP) granules, which form as a result of multivalent interactions between mRNAs and RNA binding proteins. RNA binding proteins often contain intrinsically disordered domains, which promote RNP granule assembly 
and contribute to their dynamic properties \cite{tauber2020mechanisms}. RNP granules can subcompartmentalize the cytosol for regulated colocalization or segregation of interacting proteins and RNAs \cite{langdon2018new}. Notably, 
mRNAs that are sequestered into droplet phases may be inaccessible to translation  \cite{tsang2019phosphoregulated}, reducing effective mRNA copy numbers within the cell. Although, at first consideration, reducing mRNA copy numbers would appear to increase the noise in protein numbers, here we analyze how, by reducing fluctuations in mRNA copy numbers, phase separation may paradoxically reduce noise of gene expression.
\newline 
\newline
In this paper, we analyze several biophysical scenarios in which mRNAs are segregated into distinct phases. The central and unifying assumption in all our models will be that mRNAs in the droplet phase are translationally inert, 
so that only ``free" mRNAs  are accessible to ribosomes. We first approximate the dynamics of the two mRNA state system by a random-telegraph process, and quantify noise repression with exact and asymptotic results, providing baseline results with which to compare systems in which phase-separation is incorporated. We then discuss several models that incorporate nonlinear transition rates to more accurately emulate the physics of phase separation. The models will chiefly be differentiated by the way in which phase separation affects mRNA stability. Presence of RNPs may either decrease or increase mRNA lifetimes; even when homologous RNA/protein pairs are expressed in closely related species. For example, the mRNA \textit{CLN3} is known to form phase separated droplets with a protein partner Whi3. When Whi3 is deleted in \textit{S. cerevisiae} cells, \textit{CLN3} lifetimes increase \cite{cai2013effects}, suggesting that phase separated mRNAs turnover more quickly than dilute mRNAs in cytosol. Conversely, in the genetically similar filamentous fungus \textit{Ashbya gossypii}, phase separation increases \textit{CLN3} lifetimes \cite{lee2013protein}. 
 To address both of these functions of RNP bodies, we separately consider cases where mRNA decay primarily occurs in the cytoplasm and in the dense phase. In both models, we run stochastic simulations to determine how these mechanisms influence noise in gene expression, with particular focus on gene networks with infrequent, bursty transcription. Although, in common with previous modeling \cite{klosin2020phase}, we initially consider a deterministic critical concentration of mRNA for the onset of phase separation. We then introduce a phenomenological model of time fluctuation phase separation thresholds and through simulations and analytical results, show that robust suppression of protein noise remains possible, even when the threshold mRNA concentration is allowed extensive variation. To support our theoretical findings, we perform analysis on existing smFISH data on the distribution of\textit{CLN3} mRNA transcripts within cells of the model filamentous fungus, \textit{Ashbya gossypii}. 
 
\section*{Materials and methods}
\subsection*{Analytical Methods for Quantifying Noise}

For a well-mixed, chemically reacting system consisting of memory-less reactions, the chemical master equation can be utilized to determine the evolution of the system in a probabilistic sense \cite{ash2014topics}. Consider a system in which \textit{s} different species react, and are present, at time $t$, at abundances $\bm{N} = (N_1,\dots,N_s)$. The reaction $R_{i}(\bm{N})$, $i=1,2,\dots,n$, is defined by the population reset map $\bm{N} \to \bm{N} + \bm{\eta}_{i}$, is assumed to have propensity $a_i(\bm{N})$, which can be derived from the law of mass action. The probability $P(\bm{N},t)$ that the system is in state $\bm{N}$ at time $t$ satisfies the chemical master equation \begin{equation} \label{master}
    \pdv{P(\bm{N},t)}{t} = \sum_{i = 1}^{n} \big[ a_{i}(\bm{N}-\bm{\eta}_i)P(\bm{N}-\bm{\eta}_i,t) - a_{i}(\bm{N})P(\bm{N},t) \big]
\end{equation}
This infinite set of differential equations is not easily solved, but it is relatively simple to extract the statistical moments of the probability distribution. It can be shown \cite{hespanha2005model} that for any continuously differentiable function $\psi(\bm{N},t)$, \begin{equation} \label{moment_eq}
    \frac{d \langle \psi(\bm{N},t) \rangle}{dt} = \bigg \langle \sum_{i=1}^n \big[\psi(\bm{N}+\bm{\eta_i},t)-\psi(\bm{N},t)  \big] a_{i}(\bm{N}) \bigg \rangle
\end{equation}
In particular, consider $\mu_{\bm{M}}(t) = \langle N_1^{m_1}\dots N_1^{m_1} \rangle$, which denotes the $\bm{M}$th order moment of $\bm{N}$. If all reactions $a_i$ are linear in $\bm{N}$, then \begin{equation} \label{closure}
    \frac{d \mu_{\bm{M}}(t)}{dt} = F_{\bm{M}}(\underline{\mu}(t)),
\end{equation} where $F_{\bm{M}}$ is only a function of moments of order less than or equal to $\bm{M}$. Thus for any desired moment, we can solve a closed system of differential equations to determine its time evolution.
\subsection*{Analysis of gene expression without mRNA state changes}

\begin{figure}
\begin{floatrow}
\ffigbox{%
  \includegraphics[width=1\linewidth]{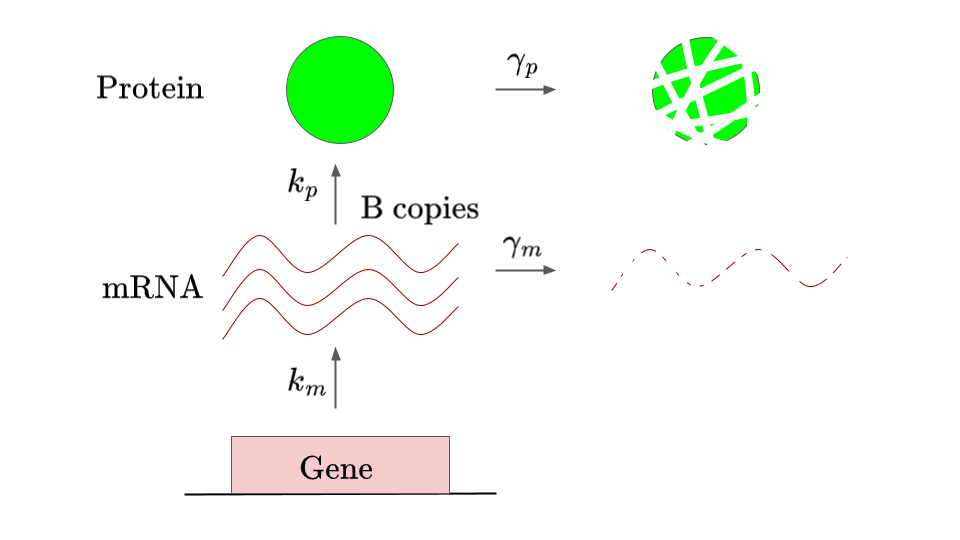}
}{%
  \caption{Schematic of gene expression without mRNA state changes. A random number B of mRNAs are transcribed with rate $k_m$; mRNAs are tranlated into proteins at rate $k_p$ and decay at rate $\gamma_m$. Protein decay rate is $\gamma_p$.}%
  \label{uged}
}

\capbtabbox{%
\scalebox{0.8}{
    \begin{tabular}{ccc}
    Event  & Population Update \tnote{*} & Propensity\\
     \midrule\midrule
Transcription  &   
$m(t) \to m(t) + B $        &  $k_m$                      \\
\midrule \\
Translation & $p(t) \to p(t)+1$ & $k_p m(t)$ \\
\midrule \\
mRNA Decay & $m(t) \to m(t) - 1$ & $\gamma_m m(t)$ \\
\midrule \\
Protein Decay & 
$p(t) \to p(t) - 1$ & $\gamma_p p(t)$ \\
\midrule \midrule\\
    \end{tabular}
}}{%
  \caption{Reaction propensities and population updates for unregulated gene expression model.}%
  \label{t1}
}
\end{floatrow}
 
\end{figure}

Let $k_m$, $k_p$, $\gamma_m$, $\gamma_p$ denote the transcription, translation, mRNA decay, and protein decay rates respectively. Roles of the kinetic constants are summarized in Fig 1 and Table 1. In addition we will assume that transcription occurs in bursts, that is, in each transcription event $B$ mRNAs are transcribed, where $B$ is a geometric random variable. By equation (\ref{moment_eq}), the moment equations of the unregulated system are, up to second order moments: 
  \begin{equation} \label{unregulated_moment}
    \begin{split}
    \dot{\left< m \right>} &= k_m \langle B \rangle - \gamma_m \left<  m \right> \\
    \dot{ \left< p \right>} &= k_p \left< m \right> - \gamma_p \left< p \right> \\
    \dot{\left< m^2 \right>} &= k_m \left(2 \langle B \rangle^2 - \langle B \rangle \right) + (\gamma_m + 2 k_m \langle B \rangle ) \left< m\right> - 2\gamma_m \left< m^2 \right> \\
    \dot{\left< p^2 \right>} &= k_p \left< m \right> + \gamma_p \left< p \right> + 2k_p \left< m p \right> - 2 \gamma_p \left< p^2\right> \\
    \dot{\left< mp \right>} &= k_p \left< m^2 \right> +k_m \langle B \rangle  \left< p \right> - (\gamma_m + \gamma_p) \left< m p\right>
    \end{split}
\end{equation}
From the first two equations, we immediately find that \begin{equation} \label{means}
    \begin{split}
        \overline{\langle m \rangle} &= \frac{k_m \langle B \rangle}{\gamma_m} \\
        \overline{\langle p \rangle} &= \frac{k_p  \overline{\langle m \rangle}}{\gamma_p},
    \end{split}
\end{equation}
where $\overline{\langle \cdot \rangle}$ denotes the steady-state ensemble average. The relative magnitude of fluctuations at equilibrium can be characterized by the quantity \begin{equation} \label{CV2}
    CV^2 = \frac{\overline{\langle p^2 \rangle} - \overline{\langle p \rangle^2}}{\overline{\langle p \rangle^2}}
\end{equation}
In this elementary gene expression model, we find that \begin{equation} \label{CV2_unregulated}
    CV^2 = \frac{1}{\overbar{\langle p \rangle}} + \frac{1 }{\left( 1 + \frac{\gamma_m}{\gamma_p} \right)} \frac{\gamma_m}{k_m}
\end{equation}
Observe that the first term in (\ref{CV2_unregulated}) represents Poisson noise; it is set only by the target protein abundance and is unaffected by any of the chemical rates. By contrast, the second term is affected by the ratios of $\frac{\gamma_m}{\gamma_p}$, and $\frac{k_m}{\gamma_m}$. Focusing on the first ratio; $\frac{\gamma_m}{\gamma_p}$ is the ratio of the protein to the mRNA lifetime. Increasing this ratio decreases $CV^2$, since transcribing short lived mRNAs into long-lived proteins smooths out the rapid fluctuations in mRNA distribution. The second ratio $\frac{k_m}{\gamma_m}$ appears in equation (\ref{means}). If $\langle B \rangle$ is held constant, then $\frac{k_m}{\gamma_m}$ is proportional to $\overline{ \langle m \rangle }$. Increasing this ratio decreases $CV^2$, since this would increase the mean mRNA abundance and thus decreases relative fluctuations in mRNAs. If $\overline{\langle m \rangle }$ is held constant, then the ratio is inversely proportional to $\langle B \rangle$. Increasing this ratio then decreases the mean burst size, decreasing the variance in burst distribution and therefore decreasing $CV^2$.
\newline

\subsubsection*{Numerical Simulation}
Our chemical reaction model is a continuous time Markov chain. Numerical simulations were performed using Gillespie's Algorithm, which provide statistically accurate trajectories for the systems \cite{gillespie1977exact}. All processes, $\bm{N}(t)$, considered in this paper are ergodic, meaning that the time average $\langle \bm{N} \rangle_T$ converges in squared mean to the ensemble average $\overline{\langle \bm{N} \rangle}$ as $T \to \infty$ Long time averages of multiple trajectories were computed to estimate statistical properties such as means and variances of state variables. \\

Reaction constants and mean molecule abundances were chosen in accordance with studies on gene expression kinetics in yeasts. In our simulations, we set $\overline{ \langle m \rangle } = 20$  and $\overline{\langle p \rangle} = 2000$. These values were chosen to be consistent with the results of Gygi et al. \cite{gygi1999correlation} who found across a large number of genes in \textit{Saccharomyces Cerevisiae} that mRNAs abundance ranged approximately from $1$-$500$ copies/cell, while protein abundance varied from $10^3$-$10^5$ copies/cell. The simulated mean abundances were specifically chosen to be small, so that relative fluctuations are large in the absence of noise suppression mechanisms. Pelechano et al. \cite{pelechano2010complete} found that for most genes in yeast, between $2$-$30$ mRNAs are transcribed per minute, which in our model represents the net transcription rate, $k_m \langle B \rangle$, the mean number of bursts per minute times the average burst size. Unless explicitly varied, we set $k_m = .05$ mRNAs/min and $\langle B \rangle = 20$, so that the net transcription rate of $1$ mRNA/min falls on the lower end of the spectrum measured by Pelechano et al. The cytoplasmic decay rate was set to $\gamma_m = .05$ mRNAs/min, so that we achieve a mean mRNA abundance $\overline{\langle m \rangle} = 20$ while still in agreement with physiologically relevant mRNA decay rates quantitated by Chia et al. \cite{chia1979half}. For all simulations, the protein decay rate was set to $\gamma_p = .02$ proteins/min, so that $t_{1/2} \approx 35$ minutes, consistent with the findings of Belle et al. \cite{belle2006quantification} on protein half-life, across 3,751 genes in \textit{Saccharomyces cerevisiae}. Finally we chose the translation rate $k_p = 2$ proteins/min, which ensures that $\overline{\langle p \rangle} \approx 2000$ in all of our simulations. These rates are summarized in Table (\ref{parameters}).
\begin{figure}
\capbtabbox{%
\scalebox{0.8}{
    \begin{tabular}[!htp]{cc}
    Parameter  & Base Value(s) \tnote{*} \\
     \midrule\midrule
$k_m$ (Burst frequency)  &   
 .05 min$^{-1}$                    \\
\midrule \\
$k_p$ (Translation rate) & 2 Proteins$(\text{min}\cdot \text{mRNA})^{-1}$  \\
\midrule \\
$\gamma_m$ (Cytoplasmic mRNA decay rate) & .05 $\text{min}^{-1}$  \\
\midrule \\
$\gamma_p$ (Protein decay rate) & 
.02 min$^{-1}$  \\
\midrule \\
$\langle B \rangle$ (Mean burst size) & 
20 \\
\midrule
$\overline{\langle m \rangle}$ (Mean mRNA abundance) & 
20 \\
\midrule
$\overline{\langle p \rangle}$ (Mean protein abundance) & 
2000  \\
\midrule \midrule\\
    \end{tabular}
}}{%
  \caption{Base reaction rates and molecular species abundances used in numerical simulations.}%
  \label{parameters}
}
\end{figure}
\subsection*{FISH Methods}
Wildtype \textit{A. gossypii} were grown in 20 ml Ashbya full media (AFM) with ampicillin 
(100 $\mu$g/ml) in a 125 ml baffled glass flask, shaking at 30°C for $\approx$ 16 hr. The cells 
were then fixed with 3.7\%  (v/v) formaldehyde for 1 hr at 37°C. After fixation, the 
cells were collected by centrifugation at 300 rpm for 5 min and washed twice with 
DEPC treated ice cold Buffer B (1.2 M sorbitol, 0.1 M potassium phosphate, pH 7.5). 
The cells were next suspended in 1 ml spheroplasting buffer (10 ml buffer B, 2 mM 
vanadyl ribonucleoside complex) and transferred to a new RNase-free 
microcentrifuge tube. The cell wall was digested by incubating the cells with 1.5 mg
Zymolase for (Sunrise Science) at 37°C for $\approx$ 40 min until cells were phase dark. 
Cells were collected by centrifuging at 2000 rpm for 2 min and washed twice with 
Buffer B. The cells were then incubated in 1 ml RNase free 70\% EtOH at 4°C for 4 
hr. Stellaris \textit{CLN3} Tamara RNA FISH probes were prepared by resuspending the 
oligonucleotide blend in 20 ul of TE buffer (10 mM TrisCl, 1 mM EDTA pH 8) to make 
a 250 $\mu$M solution. A 1:10 dilution was made of this as the working concentration to 
add to cells. After incubation at 4°C, the cells were resuspended in 1 ml wash 
solution (20× SSC, 10\% v/v deionized formamide) and allowed to reach room 
temperature. The cells were then resuspended in 100 $\mu$l hybridization buffer (1 g 
Dextran sulfate, 10 mg E. coli tRNA, 2 mM vanadyl ribonucleoside complex, 2 mg 
BSA, 20× SSC, 10\% v/v deionized formamide) with 1 $\mu$l of 25 $\mu$M probe added. The 
cells were left to incubate in the dark at 37°C overnight. The next day, cells were 
washed with 1 ml wash buffer and then resuspended in 1 ml wash buffer and 
incubated at 37°C for 30 min in the dark. The cells were then resuspended in 500 $\mu$l
wash buffer with 1 $\mu$l Hoechst (Thermo Fisher) and incubated in the dark for 15 min 
at room temperature. The cells were washed with 500 $\mu$l wash buffer with as much 
buffer removed as possible. The cells were mounted on a RNase free microscope 
slide with 20 $\mu$l mounting media (ProLong Gold antifade reagent, Invitrogen) and 
RNase free coverslip. The slide was sealed with nail polish and imaged using a 
Nikon Eclipse widefield microscope and a Plan Apo $\lambda$ 100×/1.45 oil Ph3 DM 
objective. Images were taken using phase, 405 nm, and 561 nm laser sequentially 
through a z-stack with an Andor Zyla VSC-06258 camera.

A multi-step image analysis algorithm extracted cell boundaries and mRNAs contained within cells. Steerable filters were applied to the phase contrast image stacks, to find the edges of cells in their mid-point planes, and projected to form 2D images. The outlines of cells were traced in the projected images using Adobe Illustrator software running on an Apple iPad.  From the 2D-segmentation of cells, an accurate 3D segmentation was generated by locating the optimal depth to embed the 2D mask, based on maximizing the fluorescent intensity of the 2D mask. The embedding depth of the 2D mask was allowed to vary from place to place within the mask. The mask was then swept to 3D, by skeletonizing the mask, measuring the diameter of the 2D mask at each skeleton point, and including each 3D voxel within half a diameter from each skeleton point. We detected each mRNA within the 3D segmented hyphal volume by applying a median filter and considering each local maximum as a candidate mRNA. Signal to noise ratios were calculated by applying a minimum filter with radius 0.55\,$\mu$m, to calculate the background intensity for each peak, and by dividing peak by background intensities. Spots with signal to noise ratios of less than 1.4 were discarded. Background intensities were subtracted from the image, and integrated intensities were calculated for each detected spot, as the sum of the positive background subtracted voxel intensities within 0.22$\mu$m of each detected peak. A single transcript integrated intensity was calculated for each 3D segmented hyphal volume, as the 25 percentile integrated intensity. All spots were then assigned a weight (estimated number of transcripts), by dividing them by this single transcript intensity, and rounding to the nearest integer. Simultaneously, we segmented all nuclei within the each hypha, by a two-level Otsu threshold, in which we discarded all detected objects with volumes less than 5.4$\mu$m$^3$. Our analysis included a total of 81 3D-segmented hyphae, and 1244 nuclei, of which 263 were identified as being in active \textit{CLN3} transcription.

\section*{Results}

\subsection*{Linear Stochastic Phase Separation Model}
\begin{figure}
\begin{floatrow}
\ffigbox{%
  \includegraphics[width=1\linewidth]{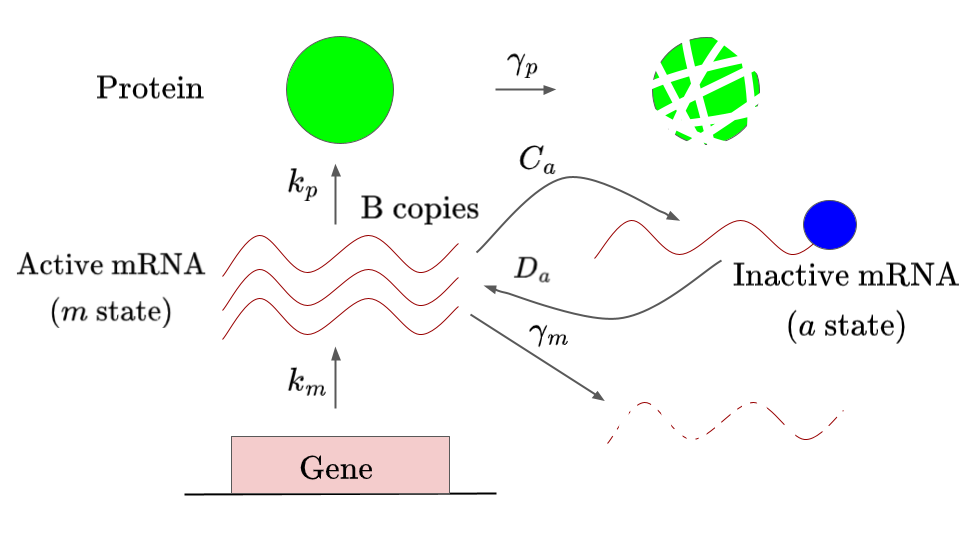}
}{%
  \caption{Schematic of Stochastic State Switching Model. In addition to the gene expression processes shown in Fig (1), mRNAs transition from active to inactive states, and back, with respective rates $C_a$, $D_a$.}%
  \label{fig:2}
}

\capbtabbox{%
\scalebox{0.8}{
     \begin{tabular}{ccc}
    Event  & Population Update \tnote{*} & Propensity\\
     \midrule\midrule
Transcription  &   
$m(t) \to m(t) + B $        &  $k_m$                      \\
\midrule \\
Translation & $p(t) \to p(t)+1$ & $k_p m(t)$ \\
\midrule \\
mRNA Decay & $m(t) \to m(t) - 1$ & $\gamma_m m(t)$ \\
\midrule \\
Protein Decay & 
$p(t) \to p(t) - 1$ & $\gamma_p p(t)$ \\
\midrule \\
Deactivation & $\begin{aligned}
    m(t) &\to m(t) - 1 \\
    a(t) &\to a(t) + 1 \\
\end{aligned}$ 
& $C_a m(t)$ \\
\midrule \\
Reactivation & $\begin{aligned}
    m(t) &\to m(t) + 1 \\
    a(t) &\to a(t) - 1 \\
\end{aligned}$ 
& $D_a a(t)$ \\
    \midrule\midrule
    \end{tabular}
}}{%
  \caption{Reactions and propensities for state switching model.}%
  \label{t3}
}
\end{floatrow}
\end{figure}

 Common to all phase separation models considered, we assume that mRNA may exist in two states outside the nucleus. In the active state, mRNA may be translated and have a decay rate of $\gamma_m$. In the inactive state, mRNAs no longer participate in translation and may either be completely stable or decay at rate $\gamma_a$.  
 To gain understanding of how mRNA state changes may affect protein copy number noise, and develop results that may be compared against nonlinear phase separation models, we first investigate a
 basic state switching model for which we can derive analytical expressions for the coefficient of variation, and compare to the $CV^2$ of the standard gene expression model (equation \ref{CV2_unregulated}). 
 The model is diagrammed in Figure (\ref{fig:2}), and the  reactions are summarized in Table (\ref{t3}). In this model, mRNAs spontaneously become inactive and the probabilities per unit time of switching between the two states are constant. While the linear kinetics of this system differ from those of phase separation, which is inherently nonlinear, its simplicity permits the computation of exact values for steady state variances, which can help demonstrate the utility of a buffer system in gene expression. We first performed stochastic simulations on this reaction network for different values of the active-to-inactive transition rate $C_a$ (Fig. 3). Qualitatively, we see that by increasing the value of $C_a$, we effectively shift mRNA copy number noise from the active mRNAs to the inactive mRNAs, resulting in smaller absolute variations in protein copy number, which is only sensitive to active mRNA fluctuations. We next performed exact calculations on this system to determine the extent to which state transitions reduce protein fluctuations.

 Similar to the single state mRNA case, the state-switching system again defines a continuous-time Markov process, so equation (\ref{moment_eq}) can be used to write down the moment equations for the first and second moments: 
\footnotesize
    \begin{equation} \label{moment_state_switch}
    \begin{split}
        \dot{\langle m \rangle } &= k_m \langle B \rangle  -\left(\gamma_m+C_a \right) \langle m \rangle  + D_a \langle a \rangle  \\
        \dot{\langle p \rangle} &= k_p \langle m \rangle - \gamma_p \langle p \rangle  \\
        \dot{\langle a \rangle } &= C_a \langle m \rangle  - D_a \langle a \rangle  \\
        \dot{\langle m^2 \rangle} &= k_m \left(2 \langle B\rangle ^2-\langle B \rangle \right) + \left(2 k_m \langle B \rangle + \gamma_m + C_a \right) \langle m \rangle  + D_a \langle a \rangle - \left(2 \gamma_m + 2C_a \right)\langle m^2 \rangle + 2 D_a \langle m a \rangle  \\
        \dot{\langle p^2 \rangle} &= k_p \langle m \rangle + \gamma_p \langle p \rangle -2\gamma_p \langle p^2 \rangle + 2 k_p \langle m p \rangle \\
        \dot{\langle a^2 \rangle} &= C_a \langle m \rangle + D_a \langle a \rangle - 2D_a \langle a^2 \rangle + 2C_a \langle ma \rangle \\
        \dot{\langle mp \rangle } &= k_m \langle B \rangle \langle p \rangle   + k_p \langle m^2 \rangle  -\left(\gamma_m + \gamma_p + C_a \right) \langle mp \rangle  + D_a \langle pa \rangle   \\
        \dot{\langle ma \rangle} &= -C_a \langle m \rangle + \left(k_m \langle B \rangle -D_a \right) \langle a \rangle  + C_a \langle m^2 \rangle + D_a \langle a^2 \rangle  - \left(\gamma_m+C_a+D_a \right) \langle ma \rangle \\
        \dot{\langle pa \rangle } &= C_a m p + k_p \langle ma \rangle - \left(\gamma_p + D_a \right) \langle pa \rangle 
        \end{split}
    \end{equation}
    \normalsize
    where dotted quantities represent time derivatives. At equilibrium, these linear equations can be solved exactly for steady-state moments. First we find that $\overline{\langle m \rangle } = \frac{k_m \langle B \rangle}{\gamma_m}$ so, surprisingly, inactivation of mRNAs does not affect their equilibrium copy number. In particular from equation (\ref{CV2}) we find that for fixed $\overline{\langle m \rangle}, \overline{\langle a \rangle}$ and $\overline{\langle p \rangle}$
    \begin{equation} \label{CV2_exact}
    \begin{split}
        CV^2 &= 
         \frac{1}{\overline{\langle p \rangle}} \\ &+ \frac{\gamma_p \left(D_a  \left(D_a\langle B \rangle  +\gamma_p \right)\overline{\langle a \rangle} + \left(D_a+\gamma_m \right)\left(D_a+\gamma_p \right) \langle B \rangle  \overline{\langle m \rangle} \right)}{\left( D_a \overline{\langle a \rangle}+\left(D_a+\gamma_m \right) \overline{\langle m \rangle} \right) \left(  D_a \gamma_p \overline{\langle a \rangle}+\left(D_a+\gamma_p \right) \left(\gamma_m+\gamma_p \right) \overline{ \langle m \rangle }\right)}
         \end{split}
    \end{equation}
\begin{figure}[t!]
\centering
\includegraphics[width=1\linewidth]{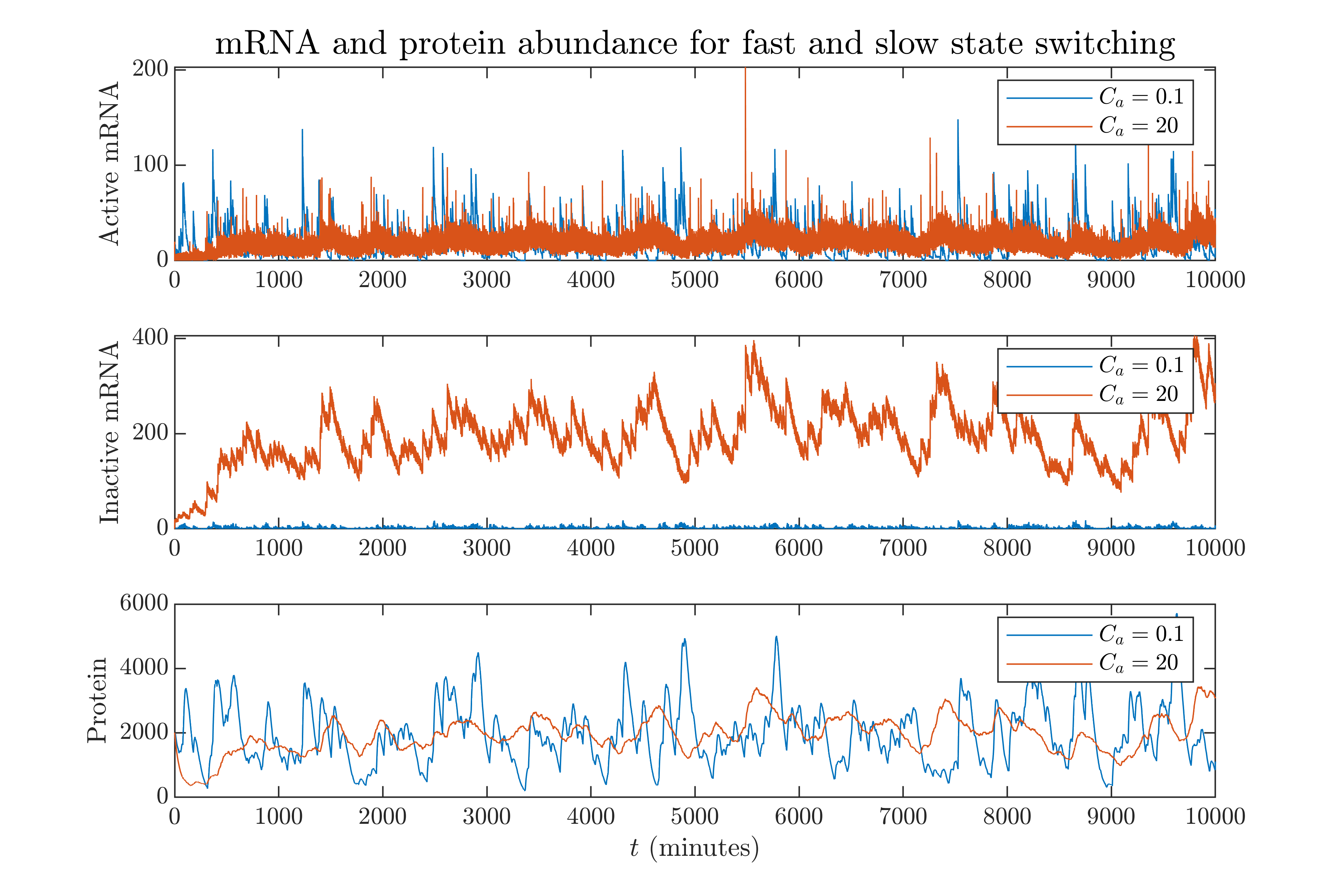}
\label{fig:3}
\caption{Two sample trajectories for model in which mRNAs can stochastically switch between active and inactive states.  By increasing the propensity for deactivation, we can decrease the magnitude of fluctuations about the common mean protein abundance of $\overline{\langle p \rangle} = 2000$. For both time series, we have $k_m = 0.05$, $k_p = 2$, $\gamma_m = 0.05$, $\gamma_p = 0.05$, $D_a = 1$, $\langle B \rangle = 20$. }
\end{figure}

\begin{figure*}[t!]
  \begin{subfigure}[t]{0.50\textwidth}
    \includegraphics[width=\linewidth]{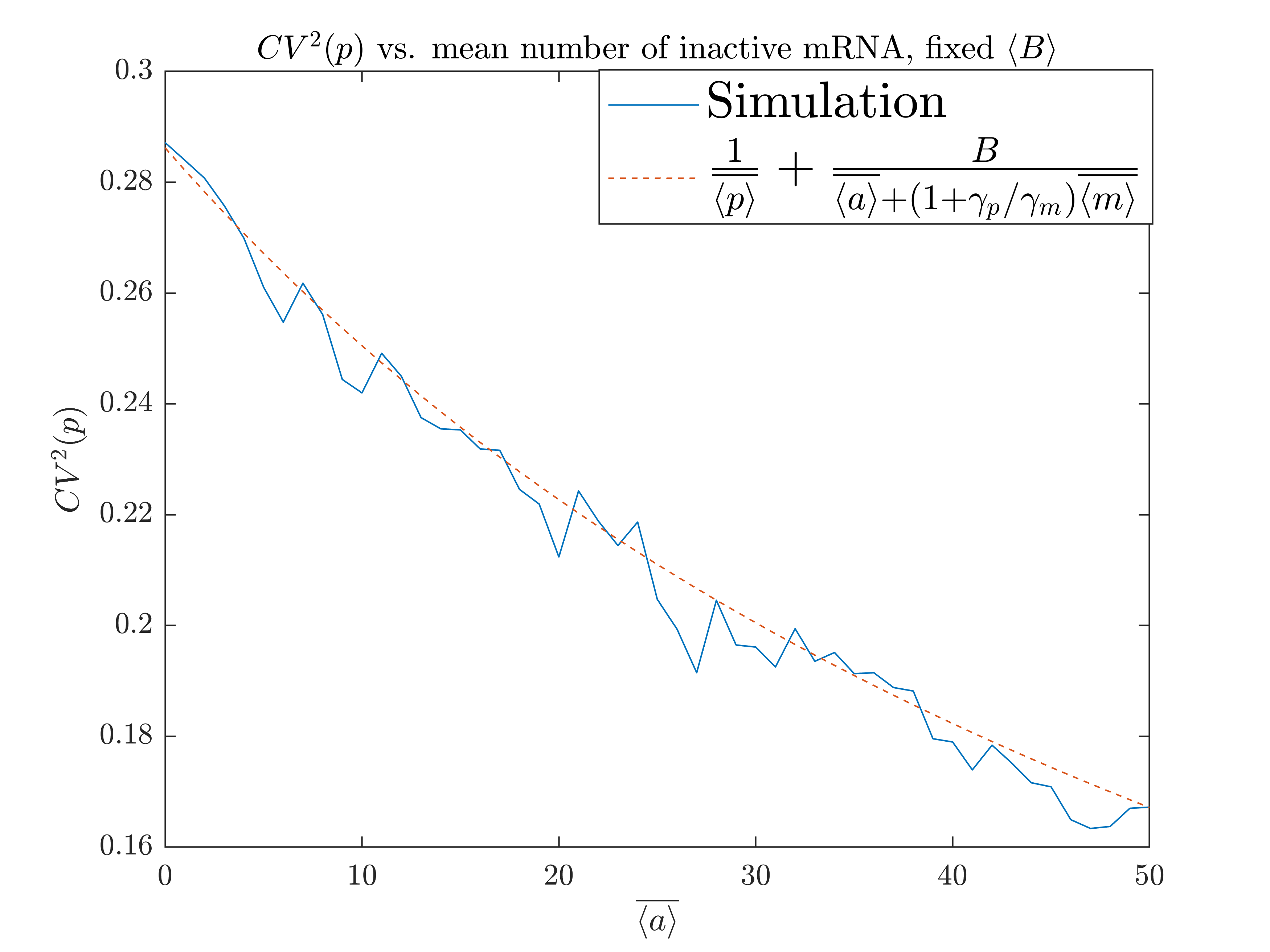}
    \caption{$k_m = 0.05$, $k_p = 2$, $\gamma_m = 0.05$, $\gamma_p = 0.02$, $\overline{\langle m \rangle} = 20$, $\overline{\langle p \rangle} = 2000$} \label{fig:4a}
  \end{subfigure}%
  \hspace*{\fill}   
  \begin{subfigure}[t]{0.5\textwidth}
    \includegraphics[width=\linewidth]{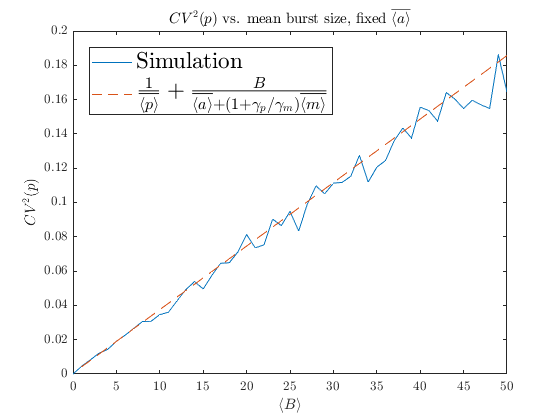}
    \caption{$k_p = 2$, $\gamma_m = 0.05$, $\gamma_p = 0.02$, $\overline{\langle m \rangle} = 20$, $\overline{\langle a \rangle} = 200$,  $\overline{\langle p \rangle} = 2000$} \label{fig:4b}
  \end{subfigure}%
  \caption{In a model in which mRNAs transition between active and inactive states, inactive mRNAs reduce protein copy number fluctuations. (a) Increasing mean inactive mRNA abundance decreases protein $CV^2$. (b) Fixing mean mRNA and protein abundance and increasing mean burst size increases protein $CV^2$. } \label{fig:4}
 \end{figure*}

If state-switching is a slow process (i.e $D_a \ll \gamma_p, \gamma_m)$, then equation (\ref{CV2_exact}) reduces to equation (\ref{CV2_unregulated}). However, in the limit that mRNA state switching is much faster than mRNA or protein decay, (i.e $D_a \gg \gamma_p, \gamma_m$), we have \begin{equation} \label{CV2_approx}
    CV^2 \approx  \frac{1}{\overbar{\langle p \rangle}} + \frac{\frac{\gamma_m}{k_m}}{\frac{\overline{\langle a \rangle}}{\overline{\langle m \rangle}}+ \left(1 + \frac{\gamma_m}{\gamma_p} \right)}
\end{equation}
Again, the first term is just the Poissonian noise of the one dimensional protein birth-death process, while the second term encodes the noise due to transcription and state switching. This term shows the same dependence upon rates $k_m/\gamma_m$ and $\gamma_m/\gamma_p$ as for unregulated translation, but also decreases monotonically with the ratio $\overline{\langle a \rangle}/\overline{\langle m \rangle}$; meaning that the inactivated mRNAs decreases protein noise, for identical transcriptional burden. We see more clearly the role of mRNA modifications when we write the transcription dependent part of $CV^2$ as \begin{equation}
    CV^2_{\text{Transcription}} = \frac{\langle B \rangle}{\overline{\langle a \rangle} + \overline{\langle m  \rangle} \left(1 + \frac{\gamma_m}{\gamma_p} \right)}
\end{equation}
and note that the term in the denominator is close to the total (active and inactive) mRNA copy number; inactivation of mRNAs buffers the system and creates an effectively larger pool of mRNAs, with effectively smaller Poissonian noise. Note however that only the factor $\overline{\langle m \rangle}$ in the denominator is multiplied by $\gamma_m/\gamma_p$, meaning that only fluctuations in transcribable mRNAs are smoothed out if the protein product is much longer lived that the mRNA. Nevertheless we observe algebraic reduction in $CV^2$ due to mRNA activation; for example, using our default parameters so $\overline{\langle m \rangle}$ and setting $\overline{\langle a \rangle} = 40$, we reduce $CV^2$ from $0.29$ to $0.19$ (Fig. \ref{fig:4a}). At the same time, the noise continues to increase if $\overline{\langle m \rangle}$ is held constant but $\langle B \rangle$ is increased - meaning that the nucleus produces mRNAs in larger but more intermittent bursts. Increasing the interval between transcriptional bursts continues to increase protein copy number fluctuations (Fig. 4b). 

\subsection*{Stochastic Nonlinear Phase Separation Model}

We now consider a nonlinear model for mRNA phase separation into ribonucleoprotein (RNP) droplets. RNP droplets form as the result of multivalent RNA-RNA, RNA-protein and protein-protein interactions. These interacting systems have been shown to exhibit concentration dependent phase transitions, conditional on both RNA and RNA-binding protein levels \cite{zhang2015rna}. To form a stochastic model of this system, we assume that the protein component of the RNP is saturating, so mRNA concentrations alone trigger the onset of droplet formation. We then define two distinct populations of mRNAs, $m$ in the dilute phase and $a$ in the droplet phase. For the stochastic simulations we use the transition rates \begin{equation} \label{Diffusion}
    \begin{split}
        C_{\text{in}} &= C \left( \sqrt{(m-m_{\text{pt}})^2+\epsilon} + (m-m_{\text{pt}}) \right)_{+} \\
        D_{\text{out}} &= D a.
    \end{split}
\end{equation}
\begin{figure}[b!]
\centering
\includegraphics[width=.5\linewidth]{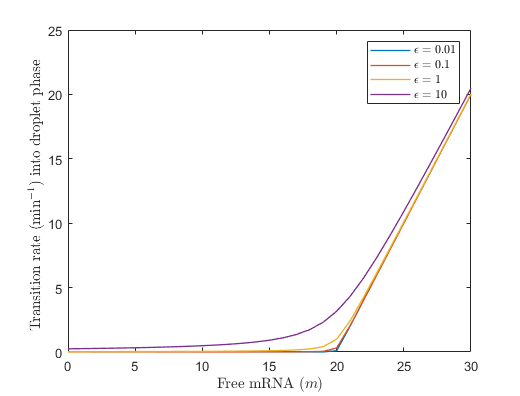}
\caption{Transition rate used to emulate a phase separation process. Droplet formation begins once $m > m_{\text{pt}} = 20$. A smoothing constant, $\epsilon$, controls the range of dilute phase concentrations over which the phase transition may occur. Here $C = 1$, $m_{pt} = 20$, and $\epsilon = 0.01,0.1,1,10$}
\label{fig:5}
\end{figure}

\begin{figure}[t!]
\centering
\begin{subfigure}[b]{0.55\textwidth}
   \includegraphics[width=1\linewidth]{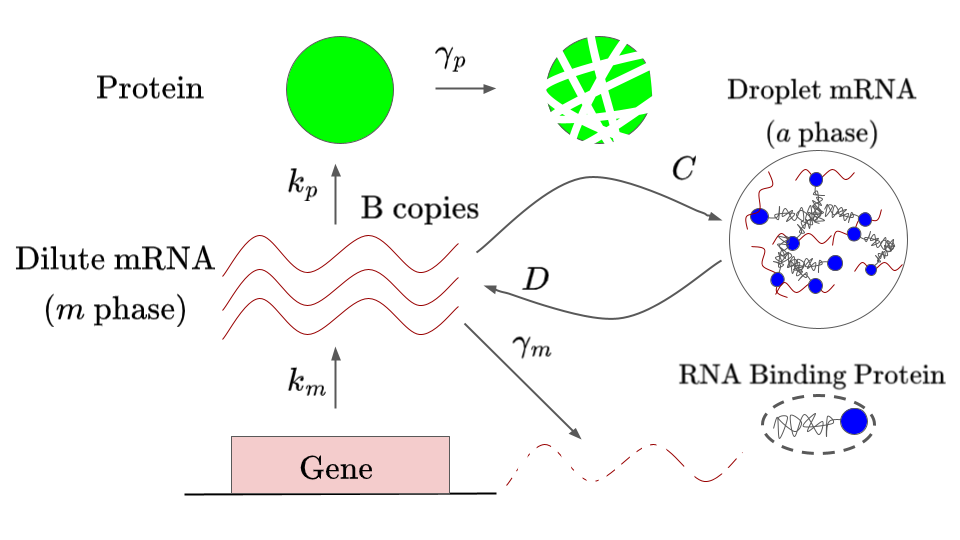}
   \caption{``Buffering" Model}
   \label{fig:6a} 
\end{subfigure}

\begin{subfigure}[b]{0.60\textwidth}
   \includegraphics[width=1\linewidth]{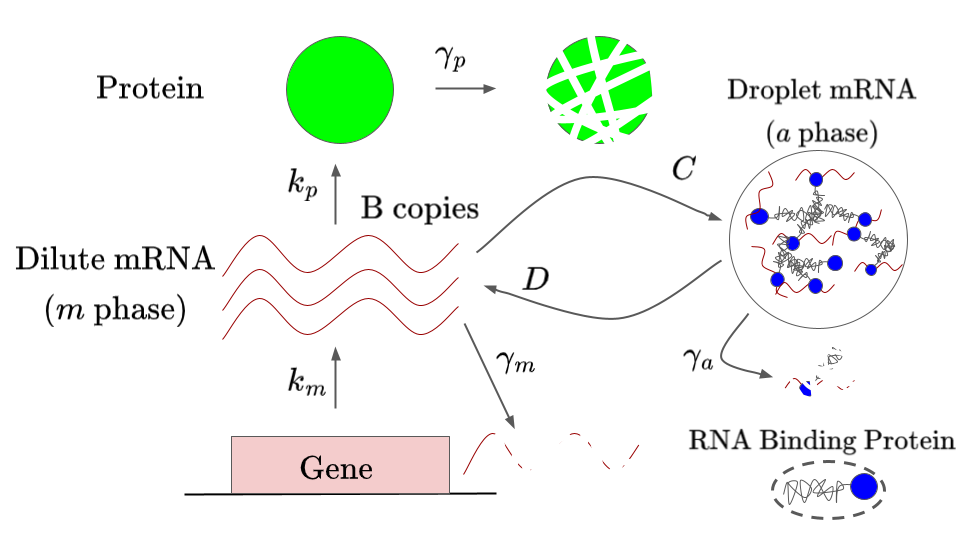}
   \caption{``Filtering" Model}
   \label{fig:6b}
\end{subfigure}

\caption[Two numerical solutions]{(a) In the buffering model, mRNA may exist in two phases; the dilute phase $m$ or the droplet phase $a$. Only mRNAs in the dilute phase may be translated into protein or decay. The propensity function $C_{\text{in}}$ was chosen so that the probability that a dilute mRNA transitions into the droplet phase is negligible if the dilute mRNA copy number $m$ is below the phase separation threshold $m_{pt}$. mRNAs within the droplet phase can transition into the dilute phase with rate $D$. (b) In the filtering model, mRNAs in the droplet phase additionally may decay with rate $\gamma_a$. We assume that $\gamma_a \gg \gamma_m$ so that decay occurs primarily in the droplet phase.  }
\label{fig:6}
\end{figure}

The phase transition of the system is characterized by the parameters $m_{\text{pt}}$ and $\epsilon$ (see Fig.  \ref{fig:5}). For $\epsilon = 0$, the transition rate for mRNAs in the dilute phase to the droplet phase reduces to \begin{equation} \label{Diffusion_approx}
    C_{in} = C \left(m - m_{\text{pt}} \right)_{+},
\end{equation}
so that transition into the droplet phase only occurs for $m > m_{\text{pt}}$. Rather than explicitly modeling the RNA binding protein, we use the parameter $\epsilon$ to influence how sharply the rate of condensate forming increases around $m \approx m_{\text{pt}}$. We test the sensitivity of our results to $\epsilon$. We take $C$ to be larger than the rate constants for translation and mRNA decay, so that state transitions dominate the dynamics of the system, occurring on the time scale of seconds in contrast to the minutes between typical transcription and decay events. Thus $\langle \overbar{m} \rangle \approx m_{\text{pt}}$, with excess mRNAs being absorbed into droplets, and the translation rate $k_p$ defined in equation (\ref{means}) can be tuned around this threshold to achieve a desired protein abundance. Using this model, we will next investigate two potential scenarios of mRNA phase separation, in the context of mRNA stability. The scenarios are shown in Fig (\ref{fig:6}). In the first scenario Fig (\ref{fig:6a}), mRNAs in the droplet phase are very long lived. We refer to this scenario as the \textit{buffer model}, since mRNAs in the droplet phase act as a reservoir that buffers dilute phase concentration of mRNAs. In a second scenario Fig (\ref{fig:6b}), mRNAs in the droplet phase decay more rapidly than dilute mRNAs. We call this scenario the \textit{filter model}, since droplets filter out excess dilute phase mRNAs.
\\

\subsubsection*{Phase Separation - Buffering Model}
\begin{figure*}[t!]
  \begin{subfigure}[t]{0.55\textwidth}
    \includegraphics[width=\linewidth]{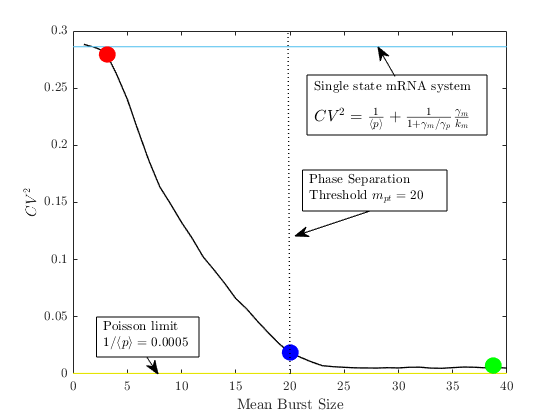}
    \caption{} \label{fig:7a}
  \end{subfigure}%
  \hspace*{\fill}   
  \begin{subfigure}[t]{0.55\textwidth}
    \includegraphics[width=\linewidth]{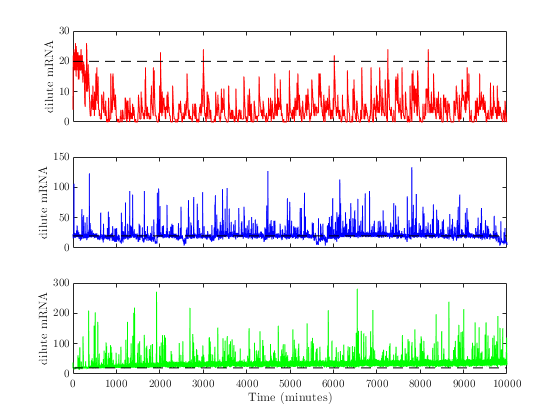}
    \caption{} \label{fig:7b}
  \end{subfigure}%
 
\caption{In the buffering model for phase separation with fixed phase separation threshold $m_{pt} = 20$, only mRNAs in the dilute phase are active. (a) For fixed bursting rate $k_m$, increasing mean burst size decreases protein $CV^2$ well below the $CV^2$ for the single mRNA state gene expression system with identical rate constants and mean protein abundance. (b) Sample trajectories of dilute mRNA population demonstrate that for a given phase separation threshold, burst intensities at or above the threshold can maintain a mean dilute mRNA abundance at $m_{pt}$ with fluctuations that are averaged out on sufficiently large timescales. Here, $k_m = 0.05$, $k_p = 2$, $\gamma_m = 0.05$, $\gamma_p = 0.02$, $C = 1$, $D = 0.01$, $\epsilon = 0.01$} \label{fig:7}
\end{figure*}

\begin{figure}[!b]
  \centering
  \includegraphics[width=0.6\linewidth]{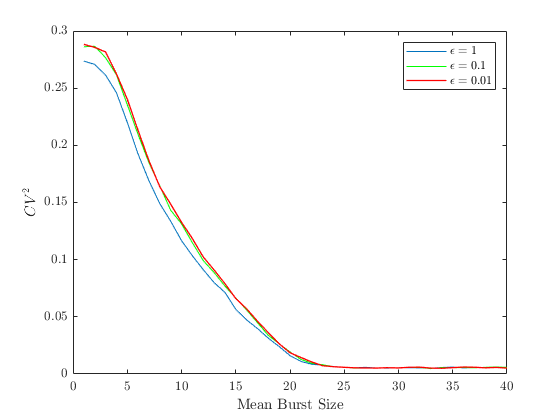}
  \caption{Figure 7a reproduced with different values of epsilon. The curves are qualitatively the same, approaching the same limiting value of $CV^2$ as the mean burst size is increased.}
\label{fig:8}
\end{figure}

For unstable dilute mRNAs, phase separation can be utilized to maintain a population of translationally inert condensate-mRNAs that become available for translation only when free mRNAs have been depleted. To test this hypothesis, we performed simulations assuming that the cytosolic mRNA decay rate $\gamma_m$ is the same order of magnitude as the protein decay rate $\gamma_p$, and that decay does not occur for mRNAs in the droplet phase, as diagrammed in Fig. 6a. We simulated protein copy number variability for different mean burst sizes (Fig 7a). The $CV^2$ of protein copy number decrease with mean burst size, plateauing when $B \approx m_{pt}$. The curve asymptotes at approximately $CV^2 = 0.006$, far below the $CV^2$ value for a system with no mRNA state changes but above the Poisson limit for protein, $\frac{1}{\overline{\langle p \rangle}} = 0.0005$, which solely measures the fluctuations due to protein birth and death.  Qualitatively, we see on examining simulated time traces (Fig 7b) that excess mRNAs transcribed in the largest bursts are quickly absorbed into the droplet phase, causing the dilute mRNA population to rapidly decrease to $m_{pt}$. Thereafter, the mRNAs in droplet phase are slowly depleted over time, reentering the dilute phase; maintaining the copy number of the dilute mRNAs close to $m_{pt}$. To test the sensitivity of the model to $\epsilon$ in equation (\ref{Diffusion}), we reproduced Figure 7a with values of $\epsilon$ ranging from $0.01$ to $1$ and found no significant qualitative differences between the $CV^2$ curves (Fig. 8). 
\\

\subsubsection*{Phase Separation: Filter Model}

  \begin{figure*}
        \centering
        \begin{subfigure}[b]{0.475\textwidth}
            \centering
            \includegraphics[width=\textwidth]{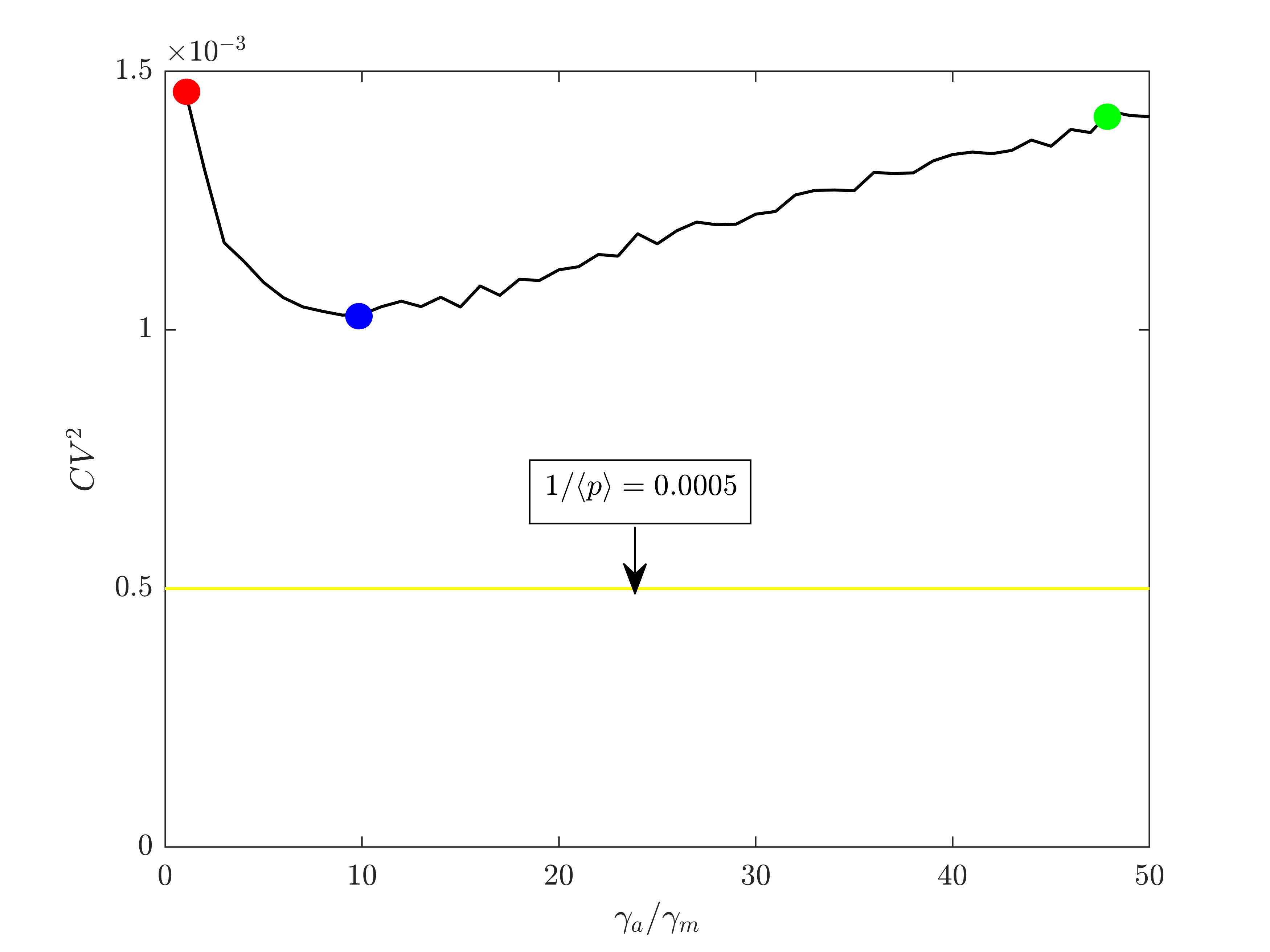}
            \caption[]%
            {{\small \textbf{Noise suppression vs. filter strength:} $k_m = 0.05$, $\gamma_m = 0.001$, $k_p = 2$, $\gamma_p = 0.02$, $\langle B \rangle = 20$, $\epsilon  = 0.01$, $C = 1$, $D = 0.01$.}}    
            \label{fig:9a}
        \end{subfigure}
        \hfill
        \begin{subfigure}[b]{0.5\textwidth}  
            \centering 
            \includegraphics[width=\textwidth]{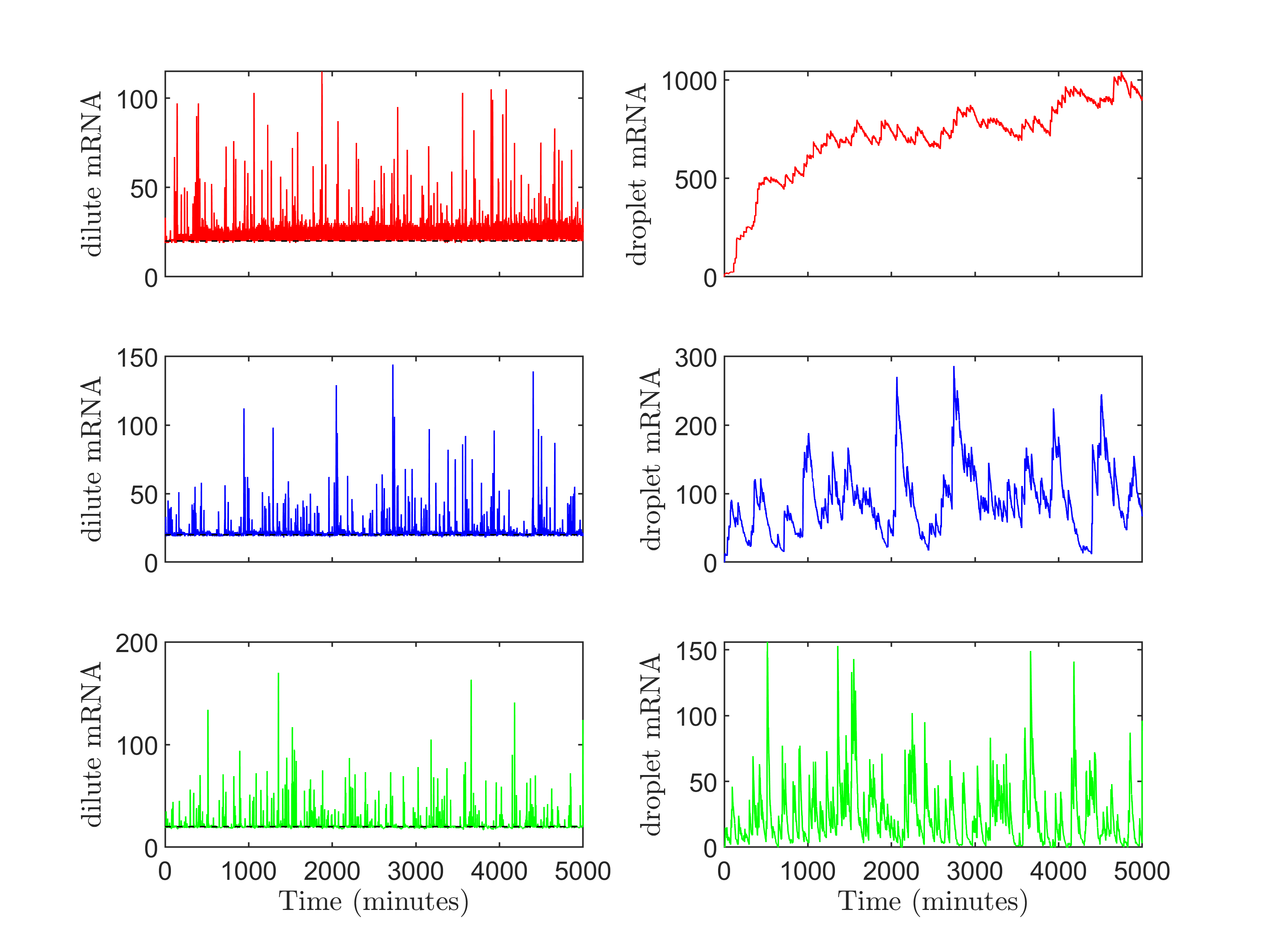}
            \caption[]%
            {{\small \textbf{Select time series of dilute and droplet mRNA:} $k_m = 0.05$, $\gamma_m = 0.001$, $k_p = 2$, $\gamma_p = 0.02$, $\langle B \rangle = 20$, $\epsilon  = 0.01$, $C = 1$, $D = 0.01$. }}    
            \label{fig:9b}
        \end{subfigure}
        \vskip\baselineskip
        \begin{subfigure}[b]{0.475\textwidth}   
            \centering 
            \includegraphics[width=\textwidth]{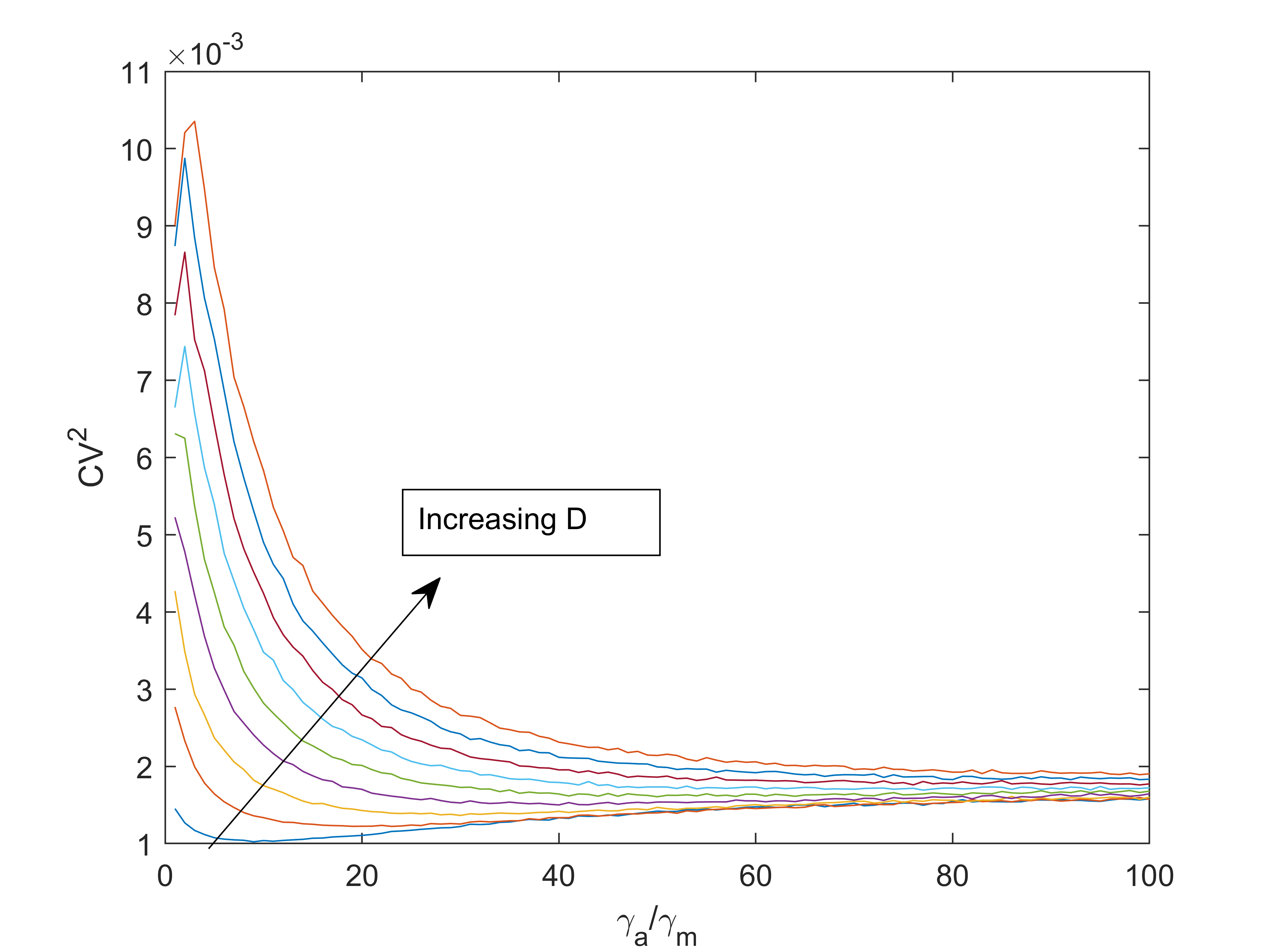}
            \caption[]%
            {{\small \textbf{Sensitivity to $D$} $k_m = 0.05$, $\gamma_m = 0.001$, $k_p = 2$, $\gamma_p = 0.02$, $\langle B \rangle = 20$, $C = 1$.}}    
            \label{fig:9c}
        \end{subfigure}
        \hfill
        \begin{subfigure}[b]{0.475\textwidth}   
            \centering 
            \includegraphics[width=\textwidth]{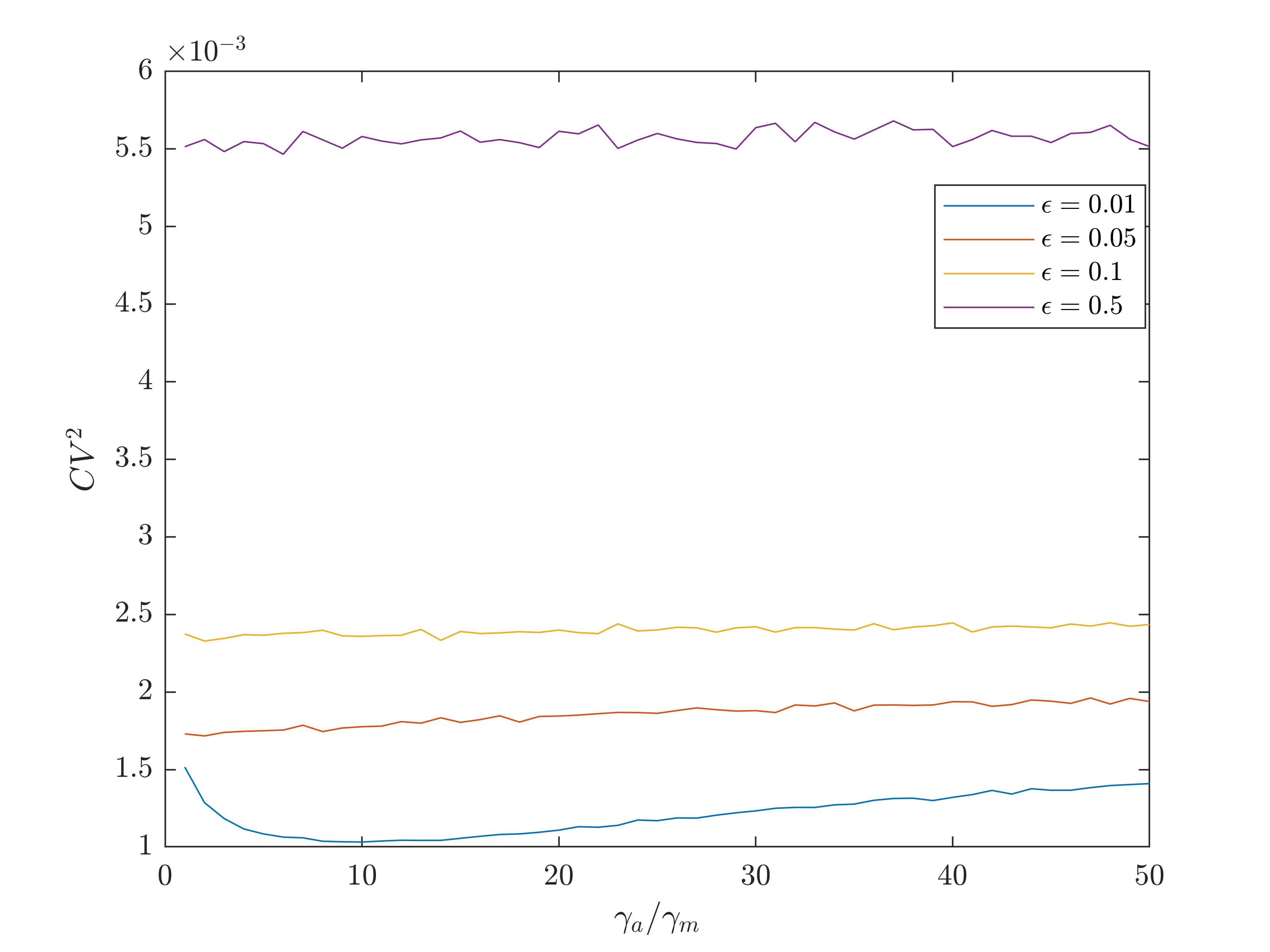}
            \caption[]%
            {{\small \textbf{Sensitivity to $\epsilon$: }$k_m = 0.05$, $\gamma_m = 0.001$, $k_p = 2$, $\gamma_p = 0.02$, $\langle B \rangle = 20$, $C = 1$, $D = 0.01$.}}    
            \label{fig:9d}
        \end{subfigure}
        \vskip\baselineskip
        \begin{subfigure}[b]{0.5\textwidth}   
            \centering 
            \includegraphics[width=\textwidth]{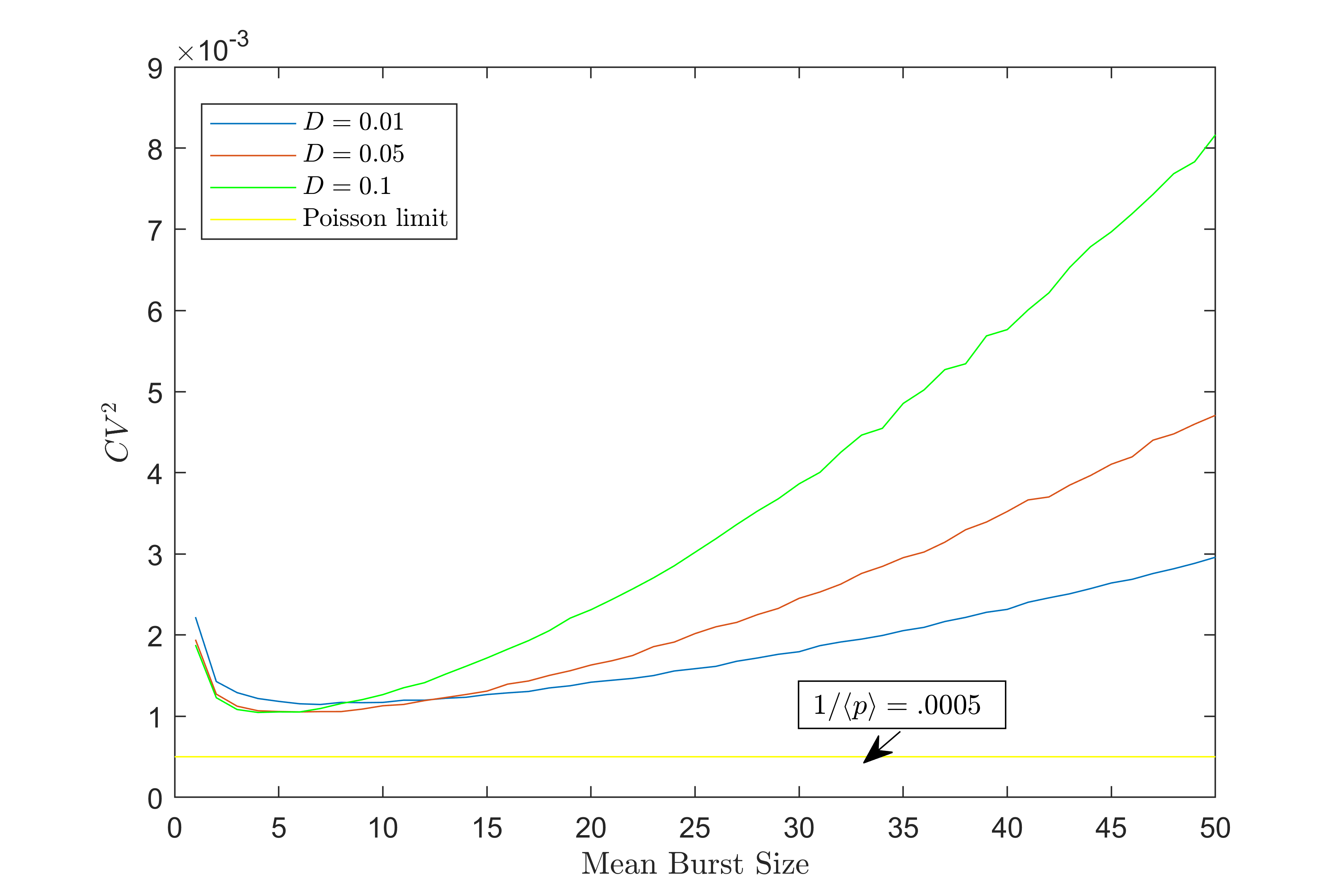}
            \caption[]%
            {{\small \textbf{Sensitivity to $\langle B \rangle$: } $k_m = 0.05$, $\gamma_m = 0.001$, $k_p = 2$, $\gamma_p = 0.02$, $\gamma_a = 0.05$, $\epsilon  = 0.01$, $C = 1$.}}    
            \label{fig:9e}
        \end{subfigure}
        \caption
        {\footnotesize In the filter model for phase separation of mRNAs, mRNAs within the droplet phase are unable to be translated and have a higher propensity for decay. (a) Varying the droplet mRNA decay rate and keeping all other parameters constant, the $CV^2$ curve exhibits a well defined minimum for $\gamma_a/\gamma_m \approx 10$. Across all values of $\gamma_a$ assayed, $CV^2$ is near the Poisson limit of protein copy number. (b) Sample trajectories of the filter model for different values of $\gamma_a$. Fluctuations in dilute mRNA copy number can be minimized by maintaining a small, but nonzero population of droplet mRNA. (c) Figure 9a replicated with different values of $D$, the diffusion rate out of the droplet phase. $CV^2$ curves shift upwards uniformly with increasing $D$, due to the reduced ability of the droplet phase to filter out excess mRNA. (d) Figure 9a reproduced for increasing values of $\epsilon$. $CV^2$ curves shift upward, but are still well below the 
$CV^2$ of the single mRNA state system.  (e) $CV^2$ exhibits a well defined minimum when varying burst size. Changing the diffusion rate out of the droplet phase alters the location, but not the magnitude of the minimum $CV^2$ of the system.} 
        \label{fig:mean and std of nets}
    \end{figure*}
 In a second scenario, mRNA stability may be decreased by interactions with RNPs \cite{cai2013effects}. Accordingly, we investigate whether decay of mRNA within protein aggregates filters cytosolic mRNA abundance and reduces fluctuations. For example, in yeasts the presence of the RNA-binding protein Whi3 reduces the half life of its target, the cyclin \textit{CLN3}, presumably by interacting with deadenylation complexes that promote turnover \cite{cai2013effects}. Thus phase separation can serve to facilitate efficient mRNA decay. We hypothesize that RNP droplets may act as a filter, causing excess mRNAs within the cytosol to be quickly degraded and recycled. To model this situation, depicted in Fig. \ref{fig:6b}, we will introduce a new reaction $a \to a - 1$ with propensity $\gamma_a a$ to our stochastic simulation, and assume that the cytoplasmic decay rate is much smaller than other decay rates, i.e $\gamma_m \ll \gamma_a, \gamma_p$. \\
We perform stochastic simulations of the filter model, varying $\gamma_a$ and holding all other parameters fixed, and measure $CV^2$ as a function of $\frac{\gamma_a}{\gamma_m}$ (Fig. 9a). We observe that $CV^2$ values are far smaller than in the system with no RNA state changes over all assayed values of $\gamma_a/\gamma_m$.  Since cytoplasmic turnover of mRNAs occurs at a much slower timescale than the other processes in the system, dilute mRNAs remain essentially fixed at the phase separation threshold, only rising when a burst occurs (Fig 9b, all left panels). Since in our model excess mRNAs rapidly transition between condensate and dilute phases, filtering allows the cytosol to quickly reach an equilibrium in which mRNAs in the dilute phase approach the phase transition concentration. This again produces $CV^2$ values that approach the Poisson limit of protein synthesis, differing only due to the small noise contributions of diffusion between states and rare burst and decay events. It is interesting to note that the $CV^2$ curve depicted in Figure 9a has a well defined minimum  $\gamma_{a}^{*} \approx 0.01$, meaning that there is an optimal strength filter for reducing noise in this parameter regime. This can be explained by the trade-off between buffering and filtering. If the filter is too weak, the mean droplet mRNA population can become much larger than the phase separation threshold, so mRNAs in droplets slowly bleed back into the cytoplasmic phase and enhance noise (Fig. \ref{fig:9b}, top row). In contrast, if the filter is too strong, the droplet population can be depleted on time scales faster than bursts, allowing cytoplasmic degradation to drive the dilute mRNAs copy number below the phase separation threshold (Fig. \ref{fig:9b}, bottom row). By maintaining a non-zero yet small droplet mRNA population, the system can benefit from both buffering and filtering effects and minimize fluctuations of dilute mRNAs about the phase separation threshold.
\par
We next investigated how the transition rate out of the droplet phase $D$, which characterizes the system's ability to buffer dilute mRNAs, influences noise profiles. We initially assumed that mRNA had a high affinity for aggregation once $m > m_{pt}$, meaning that $C \gg D$. Increasing $D$ weakens the buffering effect, and we see that the trade-off between buffering and filtering breaks down - filter strength solely determines the noise suppression ability of phase separation (Fig. 9c).  We replicated our simulations of protein-$CV^2$ as a function of $\gamma_a$, using different values of $\epsilon$, the smoothing parameter. We observe that in contrast to the buffer model (Fig. 8), the filter model is sensitive to $\epsilon$, with noise increasing with $\epsilon$. However we are still able to achieve substantial reduction in protein fluctuations compared to single mRNA state system (Fig. 9d). Finally, in Figure \ref{fig:9e}, we observe that for $\gamma_a = 0.05 \gg \gamma_m$ and different values of the droplet-to-dilute phase transition rate $D$, there exists optimal mean burst sizes $B^{*}$ where noise is minimized, balancing with the other rate parameters so that the mRNA copy number is maintained most steadily at $m_{pt}$. Increasing the diffusion rate out of the droplet phase leads to a sharper minimum that's achieved for smaller burst sizes. 

\subsection*{Stochastic Phase Separation Threshold}
 We have hitherto assumed that the protein component of the RNP droplet is saturating, which allows cytoplasmic mRNAs to be driven to and fluctuate minimally about the phase separation threshold. In live cells however, the phase separation threshold may depend on many stochastic variables, such as dynamic post-translational modifications, the concentration of other crowding proteins and the temperature of the environment \cite{alberti2017phase}. It is important to probe whether protein copy number noise is still reduced when we incorporate variabilities in the threshold for the onset of phase separation, $m_{pt}$. We model the effects of these external variables by making $m_{pt}$ subject to stochastic fluctuations. We model the fluctuations in the phase separation threshold phenomenologically as an Ornstein-Uhlenbeck process with mean $\mu$. The Ornstein-Uhlenbeck process satisifies the stochastic differential equation

\begin{equation} \label{OU_process}
    dm_{pt} = \theta (\mu - m_{pt}) + \sigma dW_t.
\end{equation}

The parameters $\theta$ and $\sigma$ characterize the strength of mean reversion and diffusivity of the process respectively. As we noted in the previous section, the dilute mRNA abundance is highly correlated with the instantaneous phase separation threshold, so we expect $\overbar{\langle m \rangle} \approx \mu$. This allows us to set $k_p = \frac{\gamma_p \overbar{\langle p \rangle} }{\mu}$ in our simulations to approximate the desired mean abundance of $\overbar{\langle p \rangle} = 2000$. When applied to the noise filter model, the fluctuations in the phase separation threshold will roughly determine the fluctuations in mRNA abundance due to the rapid phase separation dynamics. In order to demonstrate the connections between the two fluctuating variables, we performed stochastic simulations on the filter model, with $m_{pt}$ allowed to evolve according to equation (\ref{OU_process}), and computed $CV^2(p)$ for increasing values of the Orstein-Uhlenbeck noise parameter $\sigma$. To determine the extent to which these fluctuations determine noise in expression, we also compare our simulations with an exactly analyzable deterministic-stochastic system in which mRNA abundance exactly matches the varying phase separation threshold, and we assume that translation and protein decay are deterministic. In this idealized system, the random variables $m$ and $p$ satisfy the stochastic system of ODEs
\begin{equation}
    \begin{split}
        dm &= \theta (\mu - m)dt + \sigma dW_t \\
        dp &= (k_p m - \gamma_p p) dt.
    \end{split}
\end{equation}
\newline
This system can be solved in terms of $CV^2$s of the stochastic variables, i.e
\begin{equation} \label{OU_asymptotic}
\begin{split}
    CV^2(m)&= \frac{\sigma^2}{2 \theta \mu^2} \\
    CV^2(p)&= \frac{\gamma_p}{(\gamma_p+\theta)}CV^2(m).
\end{split}
\end{equation}
\\

\begin{figure}[!htbp]
  \centering
  \includegraphics[width=.5\linewidth]{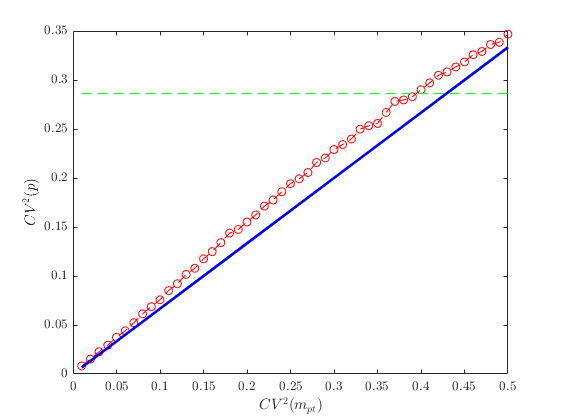}

  \label{fig:10}
  \caption{Simulations were performed on the filter model with a phase separation threshold that evolves as an Orstein-Uhlenbeck process. Here $k_m = 0.05$, $k_p = 2$, $\gamma_m = 0.001$, $\gamma_p = 0.02$, $C = 1$, $D = 0.01$, $\epsilon = 0.01$, $\langle B \rangle = 20$, $\theta = 0.01$. $CV^2$ of the protein copy number was computed for increasing values of the noise parameter $\sigma$ of the Orstein-Uhlenbeck process. Simulated $CV^2(p)$ (red) is plotted against $CV^2(m) = \frac{\sigma^2}{2 \theta \mu^2}$, with $\mu = 20$. The simulated $CV^2(p)$ curve was compared to the exact $CV^2$ of the gene expression system with deterministic translation and protein decay, and mRNA copy number driven by an Orstein-Uhlenbeck process (blue). The $CV^2$ of a single state system with the same mean mRNA and protein abundances as the red and blue systems is plotted in green.  }
\end{figure}

\begin{figure*}[t!]
  \begin{subfigure}[t]{0.45\textwidth}
    \includegraphics[width=\linewidth]{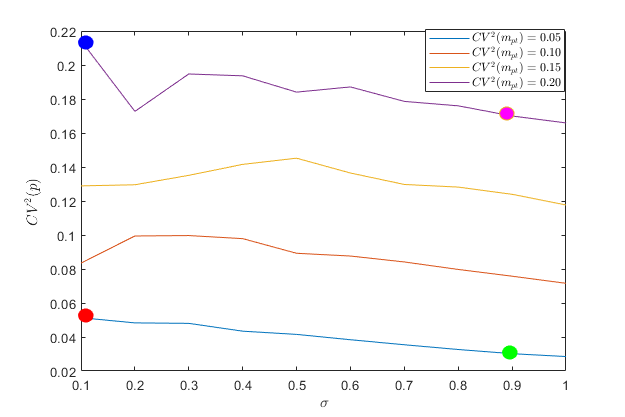}
    \caption{} \label{fig:11a}
  \end{subfigure}%
  \hspace*{\fill}   
  \begin{subfigure}[t]{0.45\textwidth}
    \includegraphics[width=\linewidth]{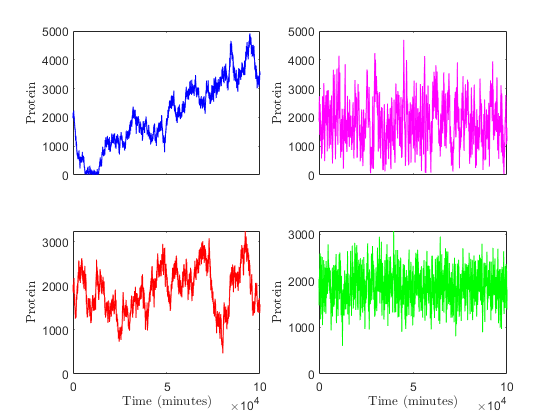}
    \caption{} \label{fig:11b}
  \end{subfigure}%
 
\caption{(a) $CV^2(p)$ values were computed for increasing values of $\sigma$ and $\theta$ so that $CV^2(m_{pt}) = \frac{\sigma^2}{2 \theta \mu^2}$ remained fixed. Curves were produced for four different values of $CV^2(m_{pt})$. (b) Sample simulations with colors corresponding to their values of $\sigma$ and $CV^2(m_{pt})$ in Fig. 11a. Here,  $k_m = 0.05$, $k_p = 2$, $\gamma_m = 0.001$, $\gamma_p = 0.02$, $\gamma_a = 0.05$, $C = 1$, $D = 0.01$,  $\epsilon = 0.01$, $\langle B \rangle = 20$} \label{fig:11}
\end{figure*}

We see in Figure 10 that both models for fluctuating phase separation thresholds introduce a similar amount of noise in expression, suggesting that stochasticity in phase separation thresholds can not be neglected in these gene expression models. However, we see that for sufficiently small fluctuations in $m_{pt}$, we can still achieve significantly reduced variation in protein abundance compared to an single mRNA state system with identical transcriptional propensity and mean species abundances. Importantly, both the exact model and analyzable approximation agree in the limit where fluctuations in $m_{pt}$ are small. The asymptotic expression (\ref{OU_asymptotic}) confirms that $CV^2(p)$ is proportional to $CV^2(m)$ - that is fluctuations in the phase separation threshold drive fluctuations in protein abundance. The prefactor in this expression, $\frac{1}{1+\theta/\gamma_p}$, decreases monotonically with the ratio of the protein lifetime to the timescale of fluctuations in $m_{pt}$, and represents the tendency of $m_{pt}$ fluctuations to be smoothed out if mRNAs are translated into long-lived proteins, since protein abundance then averages over a long history of $m_{pt}$ fluctuations. This smoothing is seen in Figure \ref{fig:11}, in which we perform simulations where we hold $CV^2(m) = \frac{\sigma^2}{2 \theta \mu^2}$ constant, and vary $\sigma^2$ and $\theta$ in proportion. We see that faster fluctuations of $m_{pt}$ are associated with smaller fluctuations in $p$, though for our simulation parameters the effects are modest; with a 10-fold increase in $\sigma$ associated with a factor $\approx 1.2$ decrease in protein $CV^2$.

\subsection*{Analysis of transcript abundances for a real RNP forming system}

\begin{figure}[t]
    \centering
    \includegraphics[width=\linewidth]{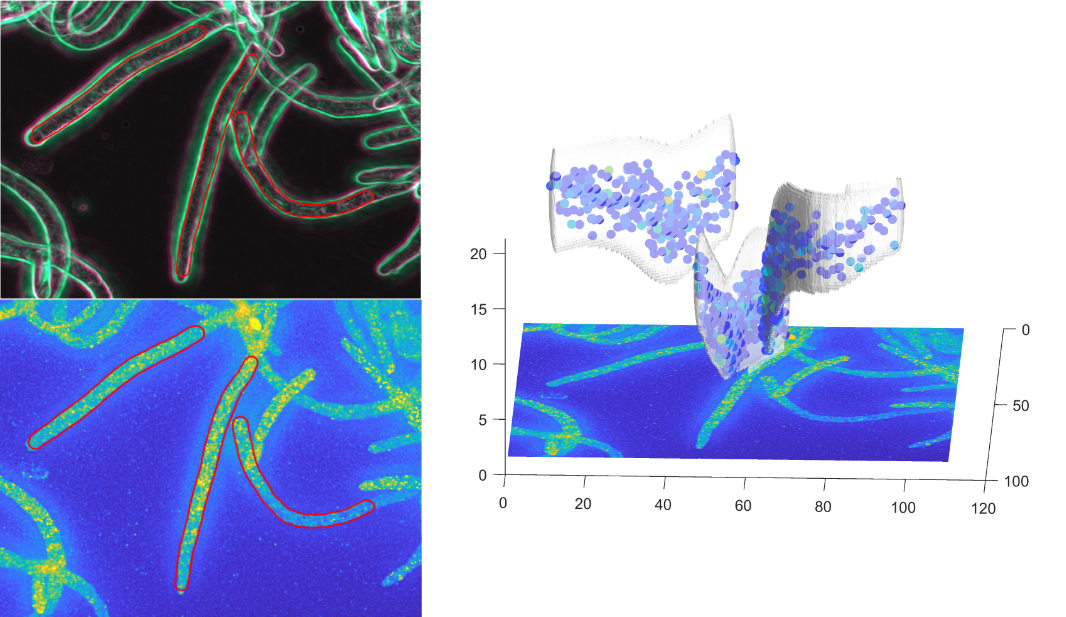}
    \caption{3D segmentation of fungal cells in \textit{Ashbya gossypii}. Top left panel: maximum intensity projection of a stack of phase contrast images of the cells, processed using steerable filters to highlight cell boundaries. Images are false colored so that edges in the middle of the $z$-stack are colored green, and edges within 3$\mu$m of the top or bottom layer are colored magenta, allowing cells that are not completely contained in the $z$-stack to be ignored. Three hyphae are shown outlined. Bottom left panel: Fluorescence image showing the \textit{CLN3} mRNA spots. Spots are shown in the 3D image color coded by the number of mRNAs that each is inferred to contain, ranging from 1 (blue) to 12 (yellow). Right panel: 3D reconstruction of the 3D cell surfaces, suspended above fluorescence image.}
    \label{fig:my_label}
\end{figure}
We tested whether our modeling could be consistent with existing data on RNP-forming systems. Distributions of \textit{CLN3} mRNA transcripts -- which are translated into the cyclin Cln3 -- were mapped in the filamentous fungus \textit{Ashbya gossypii} (see Fig 12.) using single-molecule fluorescence in-situ hybridization \cite{lee2013protein}. \textit{CLN3} was chosen because it is known to form RNP droplets with Whi3 protein, and the specificity of \textit{CLN3}-Whi3 RNP phase separation has been related to the presence of Whi3 binding sequences within the mRNA, and to polyQ-tract driven aggregation of Whi3 macromolecules \cite{zhang2015rna}.

Based on prior results about spheres of mRNA enrichment surrounding nuclei  \cite{dundon2016clustered}, we partitioned the cytoplasm (i.e. 3D segmented hyphae, subtracting their nuclei) into spheres of radius 2.5 $\mu$m centered at nuclei centroids (nuclei and parts of the sphere that extended outside of the 3D mask were excluded, so spheres did not need to be truly spherical). Within each sphere, we measured the mRNA concentration (total number of mRNA transcripts divided by neighborhood volume), as well as the mean number of mRNAs per spot. Spots with weight $> 1$ were considered to be in a condensate. We also detected mRNA transcripts within nuclei. A nucleus was assumed to be actively transcribing mRNAs if it contained mRNA transcripts.

\begin{figure}[!htb]%
    \centering
    \subfloat[]{{\includegraphics[width=.45\linewidth]{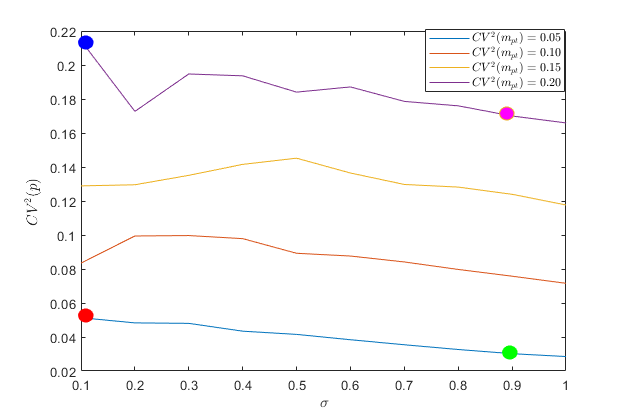} }}%
    \qquad
    \subfloat[]{{\includegraphics[width=.45\linewidth]{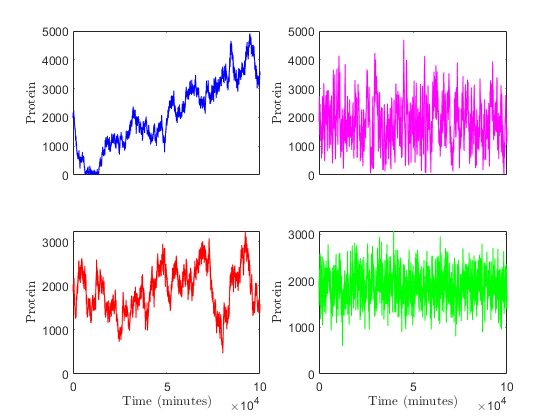} }}%
    \caption{Spheres of radius $2.5 \mu m$ were constructed around each nucleus to carve out cytoplasmic regions we refer to as \textit{nuclear neighborhoods}. Left panel: In each nuclear neighborhood, we measured the number of mRNAs divided by the neighborhood volume, and the mean spot weight. From this we formed a scatter plot, and binned the data in $40$ uniform compartments. In each bin, we measured the median mean spot weight to form the curve in blue. This curve experiences a transition at a concentration of $\approx 0.068$ mRNA,$/\mu m^3$ signifying the threshold for phase separation. With a mean neighborhood volume of $\approx 37 \mu m^3$, we estimate an average phase separation threshold of $\approx 2.5$ mRNA per neighborhood. Right panel: A histogram of the number of mRNA in actively transcribing nuclei, which we use to approximate the distribution of bursts. The mean number of transcripts synthesized in a burst was measured to be $\approx 2.4$. }%
    \label{fig:13}%
\end{figure}
In Fig. \ref{fig:13}a, binning of nuclear neighborhood data reveals evidence of a phase transition at a concentration of $\approx$ 0.068 mRNA$/\mu m^3$. The volume of a typical cytoplasmic neighborhood is $\approx 37 \mu m^3$, placing the phase transition at $\approx 2.5$ total mRNA transcripts. This number matches quite closely to the mean number of mRNA transcripts that were detected in actively transcribing nuclei, which we interpret as the burst size parameter (see Fig. \ref{fig:13}b). In reality, it's likely that \textit{CLN3} is only one of multiple RNAs contributing to the phase separation. These values reflect \textit{CLN3} thresholds, not the total RNA that may comprise the detected biomolecular condensates.

\section*{Discussion}
\begin{figure}[hb!]
\centering
\begin{subfigure}[b]{0.60\textwidth}
   \includegraphics[width=1\linewidth]{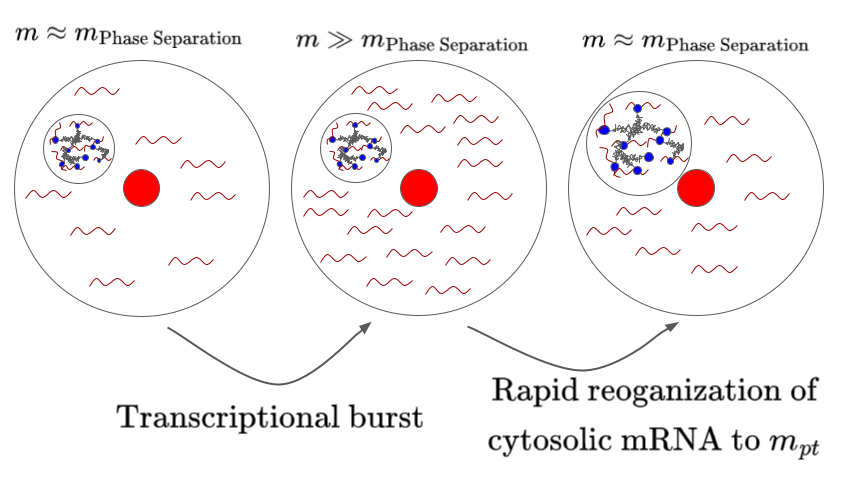}
   \caption{\centering \textit{Buffer Model}}
   \label{fig:Model3cartoon} 
\end{subfigure}

\begin{subfigure}[b]{0.75\textwidth}
   \includegraphics[width=1\linewidth]{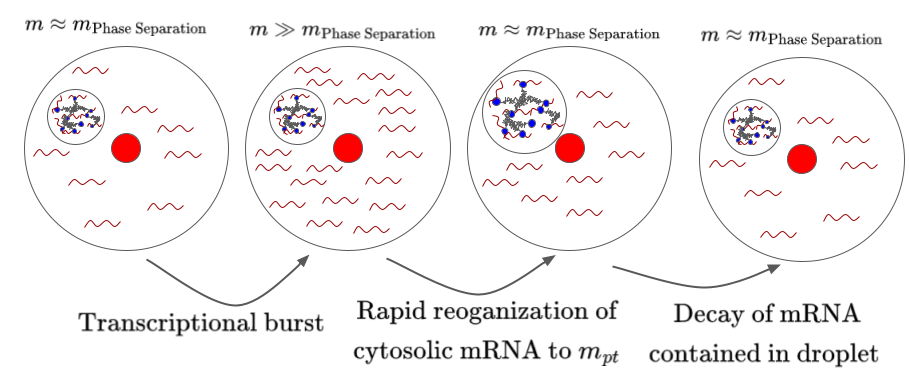}
   \caption{\centering \textit{Filter Model}}
   \label{fig:Model4cartoon}
\end{subfigure}

\caption[Two numerical solutions]{(a) Schematic of the Buffer Model. Here noise suppression is achieved through the partitioning of the cytosol between an active dilute phase and inactive droplet phase. The droplet phase acts as a reservoir for mRNAs that may have been synthesized in large bursts. So long as the net concentration of mRNAs is above the phase separation threshold, the dilute mRNA copy number remains close to the phase separation threshold with minimal fluctuations, thus reducing protein copy number noise. (b) Schematic of the Filter model. Here we assume that mRNAs are long-lived in the cytosol, and degradation primarily occurs in the droplet phase. Fluctuations due to stochastic transitions between states are reduced compared to the Buffer model, since excess mRNAs created in bursts will be quickly degraded. }
\label{fig:14}
\end{figure} 
The role of intracellular phase separation in gene expression regulation, is still an emerging field of study. Here, we developed several models to examine how phase separation of mRNAs can suppress noise in protein copy numbers. We first analyzed a linearized phase separation model and demonstrated that noise can be reduced by partitioning cytosolic mRNA into ``active" and ``inactive" states, even if the transition rates between states are constant. We then developed a phenomenological model for concentration-dependent phase separation, and used numerical simulations to quantify expression noise in two distinct biological scenarios consistent with RNP droplet function. The buffer model assumed that droplets act as a reservoir for inactive mRNAs, effectively extending their lifetime, while in the filter model mRNA decay happens primarily within the droplet phase. In both scenarios we demonstrated significant noise reduction, which was largely achieved due to the action of phase separation, which under reasonable conditions maintains the dilute mRNA copy number near the phase separation threshold (see Fig. 14). We finally introduced fluctuations into the phase separation threshold, and found that noise suppression can still be significant provided that phase separation threshold fluctuations are sufficiently rapid.

\subsection*{Efficiency in Gene Expression Regulation and Optimal Noise Suppression}
Our results indicate that through phase separation of transcripts, gene expression networks can operate efficiently with minimal signaling cues while still achieving precision in protein copy number. While infrequent transcriptional bursts can introduce a substantial amount of noise into a gene network, phase separation models show that protein abundances are largely insensitive to these large variations, so long as transitions into the droplet phase are sufficiently rapid. In fact, in the case of noise buffering, we see that in Figure \ref{fig:7a} for a fixed translation rate and phase separation threshold, that large bursts are in fact crucial in reducing noise. If bursts intensities are too small, then the network is unable to be maintained at the phase separation threshold, and transcriptional noise  becomes significant.  Due to the tradeoff between transcriptional cost and expression noise, we can imagine that selective pressures may drive burst sizes to values near the phase separation threshold, since we observe only marginal increase in noise suppression for larger bursts. In contrast, in the filter model we observe strict minima for expression noise as both a function of droplet decay rate and mean burst size. While it is difficult to predict optimal values for these parameters in live cells based on these minima due to the unquantifiable cost of bursts and selective decay machinery, these results demonstrate that kinetic parameters in phase separating systems can be tuned to effectively shift the burden of mRNA regulation from the nucleus to the cytosol.

\subsection*{Time Scale of Fluctuations}
In this paper, we were able to quantify protein abundance noise  in our models through $CV^2$, which measures variations over many realizations of identical systems. While this quantity is well represented in current literature on gene expression noise, we note here that the time scales of protein fluctuations, which do not factor into $CV^2(p)$, may also be selected for. If, for example, copy numbers fluctuate on a time scale comparable to cell cycle length, even minor deviations from the mean protein copy number could prove to be deleterious to a cell. We can observe this to some extent in Figure 3, where the small magnitude fluctuations in protein abundance are driven by slow fluctuations in in inactive mRNA. We can also see this in (\ref{fig:11}), where we introduced fluctuations into the phase separation threshold. Consider the red and magenta time series, where $\sigma_{r} < \sigma_{m}$, $\theta_{r} < \theta_{m}$. Although the $CV^2(p_r) < CV^2(p_m)$, fluctuations in the magenta protein copy number are significantly faster than the red protein, meaning that the magenta cell may produce a more favorable phenotype if relevant processes occur at a much slower time scale than its fluctuations. This suggests that the time scale of fluctuations in the phase separation threshold could prove to be an important factor in exhibiting the validity of our model, as slow variations would limit the efficacy of the buffering and filtering mechanisms. 

\subsection*{Comparison with \textit{CLN3} data}

The biological data available on the \textit{CLN3} model RNP-forming system were broadly supportive of droplets playing a role in suppressing protein fluctuations. Specifically, we found that the threshold abundance at which phase separation was evidenced to occur ($\approx$2.5 per nuclear neighborhood) was broadly comparable to the burst size inferred by counting mRNAs within nuclei (2.4). The relationship between optimal burst size and phase separation threshold is not deeply dependent upon whether condensates are filters or buffers, or equivalently whether sequestration within a condensate increases or decreases mRNA lifetimes. Indeed available experimental evidence suggests that either may be possible, including in homologous systems. For example, in data from yeast cells, which contain near-homologs of \textit{CLN3} and Whi3, suppressing Whi3 protein expression extends the lifetimes of \textit{CLN3} mRNAs, suggesting that condensates destabilize mRNAs. But data in the filamentous fungus \textit{A. gossypii} indicate that RNP droplets extend the lifetime of \textit{CLN3} mRNAs. The question of whether the droplets are functioning as buffers, or as filters, remains unanswered here, though the generality of our analysis supports noise reduction by either mechanism.

We include an additional two notes of caution here -- it is not yet possible to quantitatively measure the abundance of Cln3 proteins at the scale of individual cells, so we do not offer direct proof that Cln3 protein noise is effected by RNP formation for this system. Additionally, the main assumption of our model: that mRNAs within droplets are translation-inactive, has not yet been experimentally tested. However, RNPs are assuming an ever more central role in cell biology, and there has been a corresponding expansion, in recent years, of techniques for perturbing and measuring RNP formation kinetics in live cells \cite{langdon2018mrna}, and we expect the predictions from our mathematical model to shortly become experimentally testable.

\section*{Acknowledgments}

Financial support for this work was provided by a NIH T32 Training Grant (grant no. 5T32GM008185), and by the National Science Foundation through grant MCB-1840273.

\section*{Data Availability}
The post processed smFISH data used in this study are openly available at https://doi.org/10.5281/zenodo.7401520. Matlab scripts and figures are publicly available at: https://github.com/ajmayer95237/Noise-Paper-Files.
\nolinenumbers
\bibliography{bibliography}
\bibliographystyle{ieeetr} 

\end{document}